\documentclass[fleqn,usenatbib]{mnras}

\usepackage{newtxtext,newtxmath}

\usepackage[T1]{fontenc}

\usepackage[normalem]{ulem} 

\DeclareRobustCommand{\VAN}[3]{#2}
\let\VANthebibliography\thebibliography
\def\thebibliography{\DeclareRobustCommand{\VAN}[3]{##3}\VANthebibliography}

\usepackage{graphicx}	
\usepackage{amsmath}	
\graphicspath{{figures}}
\usepackage{url}
\newcommand{\Msun}{\,M$_\odot$}
\newcommand{\gureft}{\textsc{gureft}}
\newcommand{\molh}{H$_2$}
\newcommand{\KUV}{${\cal{K}}_\text{UV}$}

\defcitealias{Madau2014}{MD14}
\defcitealias{Bruzual2003}{BC03}
\defcitealias{Yung2024}{Y24a}
\defcitealias{Yung2024a}{Y24b}

\title[SFHs from Cosmic Dawn to EoR]{Investigating the star formation histories of galaxies from Cosmic Dawn to the Epoch of Reionization with the Santa Cruz SAM}

\author[L. Y. A. Yung et al.]{L. Y. Aaron\ Yung,$^{1}$\thanks{E-mail: l.y.aaronyung@gmail.com}
Rachel S.\ Somerville,$^{2}$
Steven L.\ Finkelstein,$^{3,4}$
and Kartheik G.\ Iyer$^{2}$
\\
$^{1}$Space Telescope Science Institute, 3700 San Martin Dr., Baltimore, MD 21218, USA\\
$^{2}$Center for Computational Astrophysics, Flatiron Institute, 162 5th Ave, New York, NY 10010, USA\\
$^{3}$Department of Astronomy, The University of Texas at Austin, Austin, TX, USA\\
$^{4}$Cosmic Frontier Center, The University of Texas at Austin, Austin, TX, USA
}

\date{Accepted XXX. Received YYY; in original form ZZZ}

\pubyear{\the\year{}}

\begin{document}
\label{firstpage}
\pagerange{\pageref{firstpage}--\pageref{lastpage}}
\maketitle

\begin{abstract}
The \textit{James Webb Space Telescope} (\textit{JWST}) has opened a new window onto galaxy evolution in the very early Universe. In this work, we leverage halo merger trees extracted from the \gureft\ dark-matter-only cosmological simulation suite together with the Santa Cruz semi-analytic model (SAM) for galaxy formation to investigate the predicted star formation histories (SFHs) of galaxies from cosmic dawn ($z\sim 14$) to the end of the Epoch of Reionization (EoR; $z\sim 6$). While we find that on average, median SFHs of galaxies across all masses are uniformly and rapidly rising over time from $14 \lesssim z \lesssim 6$ as expected, individual galaxy SFHs show a range of diverse SFHs, even for a fixed terminal mass or redshift, with bursts and mini-quenching episodes in agreement with SFHs inferred from observations. The median lookback time to form the youngest 50\% ($t_{50}$) and 90\% ($t_{90}$) of galaxies' stars decreases weakly with increasing stellar mass, and strongly with the redshift of observation. For galaxies at $z\gtrsim12$, we find typical values of $t_{50}\lesssim30$ Myr and $t_{90}\lesssim70$ Myr, a factor of $\sim$3--4 shorter than for comparable galaxies near the end of EoR ($z\sim6$). The young-star dominated nature of stellar populations in ultra-high-$z$ galaxies implies that careful modelling of young stellar populations is crucial for obtaining accurate synthetic photometry. In addition, our results have important implications for interpreting observational indicators of star formation histories and timescales.
\end{abstract}



\section{Introduction}

Since the start of its science operations, the \textit{James Webb Space Telescope} (\textit{JWST}; \citealt{Gardner2006, Gardner2023}) has pushed the observational frontier into the first few hundred million years of cosmic history, revealing an unexpectedly abundant population of distant galaxies. In particular, deep extragalactic surveys with \textit{JWST} have collectively shown that the number density of galaxies in the ultra-high-redshift universe\footnote{In this work, we use the term `ultra-high redshift' or `ultra-$z$' to refer to galaxies at $z\gtrsim10$.} far exceeds pre-\textit{JWST} expectations \citep[e.g.][]{Castellano2022, Bouwens2023, Robertson2023, Leung2023a, Finkelstein2024, Finkelstein2025, Perez-Gonzalez2025}, with many photometric candidates now confirmed by spectroscopic follow-up \citep[e.g.][]{ArrabalHaro2023, ArrabalHaro2023a, Fujimoto2023, Hsiao2024, Bunker2024, Carniani2024, Zavala2024, Zavala2024a}. 

The discovery of these luminous, early-forming galaxies has sparked vigorous debate about how and when these objects formed, and their implications for the physical processes that shape galaxy formation in a cosmological context. Their surprising luminosities and number densities have motivated efforts to infer their physical properties \citep{Adams2023, Curtis-Lake2023, Chworowsky2024, Chworowsky2026}, understand the mechanisms enabling rapid early star formation \citep{Dekel2023,Yung2024, Yung2025, Somerville2025}, and quantify the role of processes such as burstiness and enhanced light-to-mass ratios from potentially top-heavy IMFs \citep{Shen2023, Sun2023a, Gelli2024, Yung2024, Kokorev2025, Stiavelli2026}.

Over the past decade, major progress has been made in identifying the key astrophysical processes that govern galaxy formation. Modern physics-based galaxy formation models, such as cosmological hydrodynamic simulations and semi-analytic models, have become increasingly successful at reproducing a wide range of observed galaxy properties, providing a means to explore how processes across spatial and temporal scales shape galaxies over cosmic time \citep[e.g.][]{Vogelsberger2020}. However, simulating galaxies in the ultra-$z$ Universe remains extremely challenging, especially with numerical approaches. Because early galaxy populations are both less massive and rarer than their present-day counterparts, resolving them requires \emph{both} high mass resolution \emph{and} relatively large simulated volumes. Achieving these simultaneously with cosmological hydrodynamic simulations typically demands a prohibitive amount of computational resources\footnote{It is worth noting that environment-dependent zoom simulations, such as FLARES \citep{Lovell2020} and \textsc{thesan-zoom} \citep{McClymont2025}, are an effective approach for overcoming this difficulty.}.

The semi-analytic modelling approach, on the other hand, offers a powerful alternative that is computationally efficient, physically motivated, and highly modular \citep{White1991, Kauffmann1993, Cole1994, Somerville1999, Croton2006}. By adopting a set of carefully curated analytic and empirical prescriptions to represent complex processes, semi-analytic models (SAMs) can efficiently track baryonic evolution across a broad range of spatial and temporal scales. Many of these prescriptions mirror the `sub-grid' treatments employed in hydrodynamic simulations, which are often guided by similar empirical relations. The modular nature of SAMs coupled with their computational efficiency enables extensive systematic exploration of parameter space and alternative physical recipes \citep[e.g.][]{Yung2019, Somerville2025}. This approach has proven useful for generating forecasts in anticipation of deep \textit{JWST} extragalactic surveys \citep{Qin2017, Lagos2018, Dayal2019, Yung2019, Yung2019a, Hutter2021, Trinca2022} and for interpreting the resulting observations \citep{Lagos2024, Trinca2024, Yung2024, Cantarella2026, Somerville2025, Porras-Valverde2026, Dayal2025, DeLucia2026}. In this work, we use the well-established Santa Cruz SAM for galaxy formation \citep{Somerville1999, Somerville2008, Somerville2015, Popping2014}.
 
In the pre-\textit{JWST} era, the \textit{Semi-analytic forecasts for JWST and Roman} series used the Santa Cruz SAM, with free parameters calibrated only to selected $z\sim0$ constraints, to produce an extensive suite of predictions that were broadly consistent with the observational constraints available at the time, derived from deep extragalactic observations from facilities such as \textit{Hubble}, \textit{Spitzer}, and ground-based facilities. The \textit{Semi-analytic forecast} series made predictions for, and compared with available observational constraints for a broad range of statistical quantities characterizing galaxy populations. These included one-point distributions of rest-frame UV luminosities out to $z\sim10$ \citep{Yung2019} and of stellar mass and star formation rate out to $z\sim8$ \citep{Yung2019a}, ionizing photon production and intergalactic medium (IGM) constraints throughout the Epoch of Reionization \citep{Yung2020, Yung2020a, Yung2021}, as well as simulated lightcones that reproduce the spatial distribution and clustering of galaxies \citep{Yung2022, Yung2023}. Collectively, these forecasts helped motivate and inform early \textit{JWST} observing programs. 

The discovery of the abundant population of $z\gtrsim 10$ galaxies by \textit{JWST} motivates pushing these predictions to even earlier cosmic times. A practical bottleneck for extending SAM predictions into the ultra-$z$ regime has been the availability of merger trees that accurately capture halo assembly histories at very early times. State-of-the-art cosmological $N$-body simulations were generally not optimized for this regime, owing to limited mass resolution in large volumes and sparse snapshot cadence at high redshift. The \textit{Gadget at Ultrahigh Redshift with Extra-Fine Timesteps} \citep[\gureft, pronounced \textit{graft};][]{Yung2024a} suite was designed specifically to resolve halo merger trees in the ultra-$z$ Universe. \gureft\ comprises four tiered boxes that span a wide dynamic range in halo mass, with 170 snapshots stored over $6<z<40$ with a cadence of approximately one-tenth of a halo dynamical time. This naturally yields dense sampling at early times and provides the basis required to construct merger trees that accurately capture merger rates and rapid assembly. For example, the snapshot spacing is $\lesssim10$ Myr over $10 \lesssim z \lesssim 18$, comparable to the characteristic time-scales of key physical processes in the Santa Cruz SAM. Halo populations and merger-tree `branches' resolved in the individual boxes can then be `grafted' together to cover a wider dynamic range \citep{Yung2024,Somerville2025}.

The availability of reliable merger trees from \gureft\ enables the Santa Cruz SAM to be pushed into the ultra-$z$ regime out to $z\sim17$. \citet{Yung2024} leveraged this framework to quantify key uncertainties affecting comparisons between models and early \textit{JWST} measurements, including observational systematics (photometric-redshift uncertainties and field-to-field variance from limited survey areas) and theoretical uncertainties (stochasticity from short-timescale star formation variability and potential shifts in mass-to-light ratios under top-heavy IMF assumptions). Subsequently, \citet{Somerville2025} introduced a density-modulated star formation efficiency (DMSFE) framework, in which star formation in dense cloud environments proceeds with enhanced efficiency, providing a physically motivated pathway toward reconciling model predictions with the observed abundance of luminous galaxies, potentially extending agreement with \textit{JWST} constraints to $z\sim17$.

For decades, much of our empirical understanding of galaxy evolution has been built by comparing \emph{population-level} summary statistics measured across `snapshots' of the Universe, including one-point distributions (e.g., luminosity and stellar mass functions), scaling relations \citep[e.g.][]{Faber1976, Tully1977}, and spatial clustering quantified by two-point correlation functions \citep[e.g.][]{Peebles1980}. This framework has been enormously successful for establishing how the \emph{galaxy population} evolves across cosmic time. However, these statistics alone do not specify how \emph{individual} galaxies change, nor do they establish progenitor-descendant connections without additional assumptions or modelling \citep[e.g.][]{Papovich2015, Wellons2017}.

Alternatively, a galaxy’s star formation history (SFH), which describes the evolution of its star formation rate over cosmic time, captures the timing, duration, and intensity of star-forming events, providing a ``fossil record'' of the combined formation history of all of its progenitor galaxies. In principle, galaxy spectral energy distributions (SED) encode this rich collection of information about the pathways through which galaxies assemble their stellar mass across cosmic time \citep{Walcher2011, Conroy2013, Madau2014}. There has been a major effort over the past several decades to extract physical properties of galaxies, including stellar masses, star formation timescales, and even full star formation histories, by fitting modelled SEDs or SED templates to high-resolution galaxy SEDs \citep[e.g.][]{Chevallard2016, Iyer2017, Carnall2018, Boquien2019, Leja2017, Leja2019, Johnson2021, Wang2025}. 
Physically motivated SFHs have also been used to interpret individual galaxies at the redshift frontier, such as GS-z14-0 at $z=14.32$, where the resulting synthetic spectrum is used to interpret its observed properties \citep{Ferrara2024}.

However, observed SEDs depend not only on the distribution of stellar ages represented by the composite stellar population within a galaxy, but also on the stellar metallicities, nebular emission, and in some cases, radiation from an accreting black hole. Interstellar dust can also dramatically modify observed SEDs in a complex and difficult to model manner (see \citealt{Iyer2026} and references therein for an in-depth discussion). Disentangling these degenerate effects remains a difficult problem. 

Within Bayesian inference-based approaches to SED fitting, assumptions on the prior of galaxy star formation histories can dominate systematic uncertainties in the recovery of stellar masses and SFH \citep[e.g.][]{Carnall2019, Leja2019a, Lower2020, Jain2023}. Traditionally, it was common to adopt 
parametric SFHs such as exponentially declining models \citep[`$\tau$' models;][]{Papovich2001}, delayed-$\tau$ models \citep{Lee2010, Pacifici2012, Leja2019}, and log-normal forms \citep{Diemer2017, Cohn2018}. These parameterizations are motivated by idealized one-zone regulator/consumption pictures in which the star formation rate declines as the cold-gas reservoir is depleted. Such parameterizations have been tested against low- and intermediate-redshift datasets with high-quality photometry and spectroscopy (e.g. SDSS and GAMA at $z\sim0$ \citep{Gladders2013, Abramson2015, Carnall2019}, COSMOS/UltraVISTA and CANDELS/3D--HST at $z\lesssim3$ \citep{Muzzin2013, Muzzin2013a, Skelton2014, Momcheva2016}). However, these simple parametric forms fail to capture burstiness, rapid quenching, rejuvenation, or multi-component SFHs, and impose strong and potentially unphysical priors on the time dependence of star formation, which in particular may not be appropriate or representative of galaxies in the high redshift Universe.  
As a result, they can bias inferred stellar ages, stellar masses, and instantaneous SFRs, and can underestimate uncertainties when the true SFH differs substantially from the assumed functional form \citep{Lower2020}. Non-parametric SFH, adopted in some recent SED fitting codes (e.g. dense basis \citep{Iyer2017}, \textsc{prospector} \citep{Johnson2021}, BAGPIPES \citep{Carnall2018}) provide more flexible SFH descriptions, but their application at the redshift frontier remains challenging, particularly at ultra-high-$z$, where in many cases only sparse broadband photometric data are available, signal-to-noise is low, and coverage is often limited to the rest-frame UV part of the SED. In addition, strong rest-frame optical nebular lines can contaminate broadband fluxes, exacerbating degeneracies.

Theoretical simulations can provide guidance on physically informed priors for star formation histories at different cosmic epochs, potentially reducing some of these uncertainties. 
Previous studies have extracted SFHs from cosmological hydrodynamic simulations \citep{Furlong2015, Finlator2007, Finlator2011, Sparre2015, Sparre2017, Tacchella2016, Donnari2019, Wright2019, Iyer2020}, from semi-analytic models \citep{Pacifici2012, Pacifici2016, Shamshiri2015, Henriques2015,Iyer2020, Iyer2025, Legrand2021}, and from empirical models embedded in cosmological halo merger trees \citep{Behroozi2013c, Behroozi2019, Moster2013, Moster2018, Rodriguez-Puebla2025}. These approaches can capture the influence of a wide range of physical processes---including stellar and AGN feedback and merger-induced starbursts---and thereby yield detailed SFH predictions. However, none of these studies have presented results targetted at the ultra-high redshift Universe. During this epoch, we expect galaxy growth to be extremely rapid, and there is evidence that star formation is highly stochastic. Characterizations of SFH for lower redshift galaxy populations may not be representative of these higher redshift objects. 

In this work, we present new predictions by coupling the Santa Cruz SAM with \gureft\ merger trees, and we conduct an in-depth investigation of the resulting star formation histories for galaxies from Cosmic Dawn through the end of the Epoch of Reionization ($17\gtrsim z \gtrsim6$). We find that galaxies at these epochs have rapidly rising SFR towards later times, resulting in highly young-star dominated stellar populations. 
We also present updated photometry predictions and UV luminosity functions after carefully accounting for the contributions from young stellar populations, which were not properly treated in \citet{Yung2024}. We characterize the timescales over which galaxies form their stars, and present predictions for star formation rates averaged over different timescales, which can be probed with existing and upcoming observations. 

The structure of this paper is as follows. In Section \ref{sec:models}, we provide a brief overview of the Santa Cruz semi-analytic models, the \gureft\ simulation suite, and the coupling of simulated SFHs and synthetic SEDs. In Section \ref{sec:results}, we present our main results on star formation histories and the timescales over which galaxies formed their stars. We discuss the implications of our results in Section \ref{sec:discussion}, and summarize and conclude in Section \ref{sec:summary}.


\section{Simulation and modelling framework}
\label{sec:models}

In this section, we provide a concise overview of the key components of this work: the Santa Cruz semi-analytic model for galaxy formation (Section~\ref{sec:sc-sam}), the \gureft\ cosmological $N$-body simulation suite (Section~\ref{sec:gureft}), and our construction of synthetic composite SEDs by interfacing SAM-predicted SFHs with \textsc{bpass} Simple Stellar Population (SSP) models (Section~\ref{sec:sfhist_sed}).
Throughout this work, we adopt cosmological parameters consistent with \citet{PlanckCollaboration2016}: $\Omega_m = 0.307$, $\Omega_\Lambda = 0.693$, $H_0 = 67.8$ km s$^{-1}$ Mpc$^{-1}$, $\sigma_8 = 0.823$, and $n_s = 0.96$. These values are consistent with those adopted in \gureft\ \citep{Yung2024a} and the VSMDPL simulation from the MultiDark suite \citep{Klypin2016}. All magnitudes are expressed in the AB system \citep{Oke1983}, and we assume a Chabrier stellar initial mass function \citep[IMF;][]{Chabrier2003}. Unless otherwise specified, all logarithms are base 10.


\subsection{The Santa Cruz semi-analytic model}
\label{sec:sc-sam}

The semi-analytic model (SAM) developed by the Santa Cruz group, commonly referred to as the Santa Cruz SAM, is a versatile galaxy formation framework that incorporates a set of carefully curated physical processes to simulate the formation and evolution of galaxies within the scaffolding of dark matter halo merger trees \citep{Somerville1999, Somerville2008, Somerville2015, Somerville2021}. This modelling framework includes many standard components that are commonly found in other SAMs and cosmological hydrodynamic simulations, such as cosmological gas accretion and atomic cooling, suppression of gas accretion after reionization by the intergalactic UV background, star formation and stellar-driven winds, chemical evolution, black hole feedback, and mergers. We refer the reader to the papers above for a full description of the model components and to \citet{Yung2022} for a flowchart of the model's internal workflow.

The specific version of the model adopted in this work is identical to that used in \citet{Yung2024}, and is configured to utilize an \molh-based star formation relation \citep{Bigiel2008} and a multi-phase gas partitioning scheme \citep{Gnedin2011}, as introduced and implemented by \citet{Somerville2015}. The gas partitioning recipe divides the cold gas disc into atomic, molecular, and ionized components using fitting functions based on numerical hydrodynamic simulations \citep{Gnedin2011}. In a similar spirit to the Kennicutt--Schmidt star formation (SF) relation, in which the SFR is proportional to the surface density of cold gas (e.g. \citealt{Kennicutt1998}), the \molh-based SF relation assumes that the surface density of star formation scales with the surface density of molecular hydrogen (e.g. $\Sigma_\text{SFR} \propto \Sigma_{\text{\molh}}^{\alpha}$), where the slope $\alpha$ steepens from 1 to 2 above a critical molecular gas surface density, $\Sigma_{\text{\molh,crit}}$. This steepening is motivated by observations \citep{Sharon2013, Rawle2014, Hodge2015, Tacconi2018} as well as theory \citep{Ostriker2010}.

The Santa Cruz SAM also incorporates stellar- and AGN-driven feedback. Stellar feedback is implemented as large-scale outflows driven by massive stars and supernovae, with a mass-loading factor given by
\begin{equation}
\dot{m}_\text{out} = \epsilon_\text{SN}
\left(\frac{V_0}{V_c}\right)^{\alpha_\text{rh}}
\dot{m}_* ,
\end{equation}
where $V_c$ is galaxy circular velocity and $V_0=200$ km s$^{-1}$, and $\epsilon_\text{SN}$ and $\alpha_\text{rh}$ are calibrated free parameters. Ejected gas may later be re-accreted, allowing the feedback cycle to contribute to time-variable SFHs. Black holes are seeded in top-level haloes with $M_\text{BH}=10^4$ \Msun\ and grow through both merger/disc-instability-driven accretion from the ISM and Bondi-Hoyle accretion from the hot halo \citep{Somerville2008, Hirschmann2012}. The corresponding AGN feedback modes drive cold-gas ejection (through a radiation pressure driven wind) and hot-halo heating (via relativistic jets), respectively. We note that AGN feedback has no noticeable direct impact on the high-redshift results presented in this work, but mention it here because it is part of the globally calibrated SAM framework. We refer the reader to \citet{Yung2019} for a systematic exploration of how varying $\alpha_\text{rh}$ affects the predicted UV luminosity functions.

This model configuration has a well-established track record of reproducing a wide variety of observed galaxy population statistics, including rest-frame UV luminosity functions \citep{Yung2019, Yung2024} and various one-point distribution functions of physical properties \citep{Somerville2015, Yung2019a} at $4 < z \lesssim 10$, two-point auto-correlation functions \citep{Yung2022, Yung2023}, and constraints on the ionizing photon production rate and intergalactic medium (IGM) reionization \citep{Yung2020, Yung2020a}.

The galaxy formation modelling framework also includes mechanisms that contribute to episodic star formation activity, such as galaxy--galaxy mergers and merger-induced starbursts. Following a galaxy-galaxy merger, the star formation efficiency is enhanced for a time $\tau_{\rm burst}$, where this timescale is a function of the galaxy circular velocity, gas fraction, and redshift, based on hydrodynamic simulations of idealized binary mergers \citep[see][for details]{Somerville2008}. The pre-existing stellar populations from both progenitor galaxies are also added together. Episodic star formation activity, including quenching, also arises from the feedback cycle. Gas is ejected from the interstellar medium by stellar and AGN driven winds. Gas ejected by AGN is assumed to leave the halo forever and is not re-accreted, however, gas ejected by stellar driven winds can be re-accreted into the halo and then cool and flow into the ISM. Thus the model produces complex star formation histories with stochasticity on a variety of timescales, corresponding to different physical processes (e.g. merger-triggered bursts vs. gas re-accretion) \citep{Iyer2020}. However, the model is missing physical processes that may lead to additional shorter timescale burstyness, such as star formation and early stellar feedback in a clumpy ISM (see e.g. Figure~4 of \citealp{Pandya2020}). 

The model parameters are calibrated as outlined in \citet{Gabrielpillai2022}, such that the SAM outputs match the observed $z\sim0$ stellar mass function \citep{Baldry2012, Bernardi2013, Moustakas2013}, stellar-to-halo mass ratio \citep{Rodriguez-Puebla2017}, cold ISM gas fraction versus stellar mass \citep{Calette2018, Catinella2018}, stellar mass-metallicity relation \citep{Gallazzi2005, Kirby2011}, and the $M_\text{BH}$--$M_\text{bulge}$ relation \citep{Kormendy2013, McConnell2013}. These physical parameters are not re-\textit{`tuned'} at higher redshift, and we use identical parameter values in this study. Note that these parameters are slightly different from those used in the \textit{Semi-analytic forecasts for JWST} paper series \citep{Yung2019, Yung2022}, due to the switch from Extended Press-Schechter based merger trees to $N$-body-based trees.

We also note that this model configuration, like the majority of pre-JWST launch semi-analytic models and hydrodynamic simulations, broadly reproduces observed galaxy population properties from the local Universe back to about $z\sim 10$, but underproduces UV luminous galaxies at higher redshifts relative to the populations recently discovered by JWST \citep[see][and Fig.~\ref{fig:UVLF_8panels_compY24}-\ref{fig:UVLF_8panels_ext} of this work.]{Yung2024,Somerville2025}. The qualitative results of this work would remain unchanged if the updated Density Modulated Star Formation Efficiency scenario proposed by \citet{Somerville2025} were adopted.


\subsection{The \gureft\ simulation suite}
\label{sec:gureft}

Dark matter halo merger trees \citep[e.g.][]{Lacey1994, Somerville1999a} encode the hierarchical growth of structure and provide the backbone for the semi-analytic modelling approach. To capture the rapid build-up of halos in the ultra-$z$ universe, we adopt merger trees extracted from the \textit{Gadget at Ultrahigh Redshift with Extra-Fine Timesteps} (\gureft; pronounced \textit{graft}) suite of dark-matter-only cosmological simulations \citep{Yung2024a}. \gureft\ was designed specifically for the high- to ultra-high-redshift regime, where halos grow and merge on very short timescales and where the cadence of stored snapshot outputs in conventional simulation suites can limit the fidelity of extracted merger trees.

The \gureft\ suite consists of four $1024^3$-particle volumes, \gureft-05, \gureft-15, \gureft-35, and \gureft-90, with box sizes of 5, 15, 35, and 90~Mpc~$h^{-1}$ on a side, respectively, which correspond to 7.4, 22.1, 53.6, and 132 comoving Mpc for our adopted cosmology. These boxes adopt dark matter particle masses of $M_\text{DM} = 1.5\times10^{4}$, $4.0\times10^{5}$, $5.0\times10^{6}$, and $8.5\times10^{7}\,{\rm M_\odot}$, respectively. The tiered-box design is chosen to provide a wide dynamic range: the smaller volumes resolve low-mass progenitor halos relevant to the earliest stages of galaxy formation, while the larger volumes provide statistically robust samples of rarer, more massive halos that host the galaxies accessible to current deep surveys.

A key feature of \gureft\ is the exceptionally high cadence of stored outputs. For each simulation volume, 170 snapshots are stored over $40 \gtrsim z \gtrsim 6$, with snapshot spacing set to approximately one-tenth of the halo dynamical time at the output redshift. 
This choice naturally yields denser sampling at earlier times, which is essential for reconstructing accurate merger trees in the regime where halo growth and merger rates evolve rapidly. 
In the redshift range most relevant to this work, the snapshot spacing is typically $\sim$5--10~Myr, substantially finer than in most publicly available cosmological simulations. 

Dark matter halos in \gureft\ are identified with the seven-dimensional phase-space halo finder \textsc{rockstar} \citep{Behroozi2013a} and linked across snapshots with \textsc{consistent-trees} \citep{Behroozi2013b}, adopting the virial mass definition of \citet{Bryan1998}. 
The combination of wide dynamic range and dense snapshot cadence yields merger trees that are well suited for semi-analytic modelling in the ultra-$z$ regime.

In this work, we couple the Santa Cruz SAM to \gureft\ merger trees to generate predictions for galaxy star formation histories at high to ultra-high redshift.
Where a larger simulated volume is required (e.g. to better sample the rare high-mass tail of dark matter halos), we additionally make use of merger trees extracted from the Very Small MultiDark Planck (VSMDPL) simulation from the MultiDark suite \citep{Klypin2016}. VSMDPL has a box size of 236 comoving Mpc (a volume $\sim$6 times larger than \gureft-90) and a dark matter particle (DM particle) mass of $M_{\rm DM}=9.1\times10^{6}\,{\rm M_\odot}$, comparable to that of \gureft-35. Halo catalogues and merger trees were extracted using the same tools and methods used for the \gureft\ suite.


\begin{figure*}
    \hspace*{-0.25cm}
    \includegraphics[width=2.12\columnwidth]{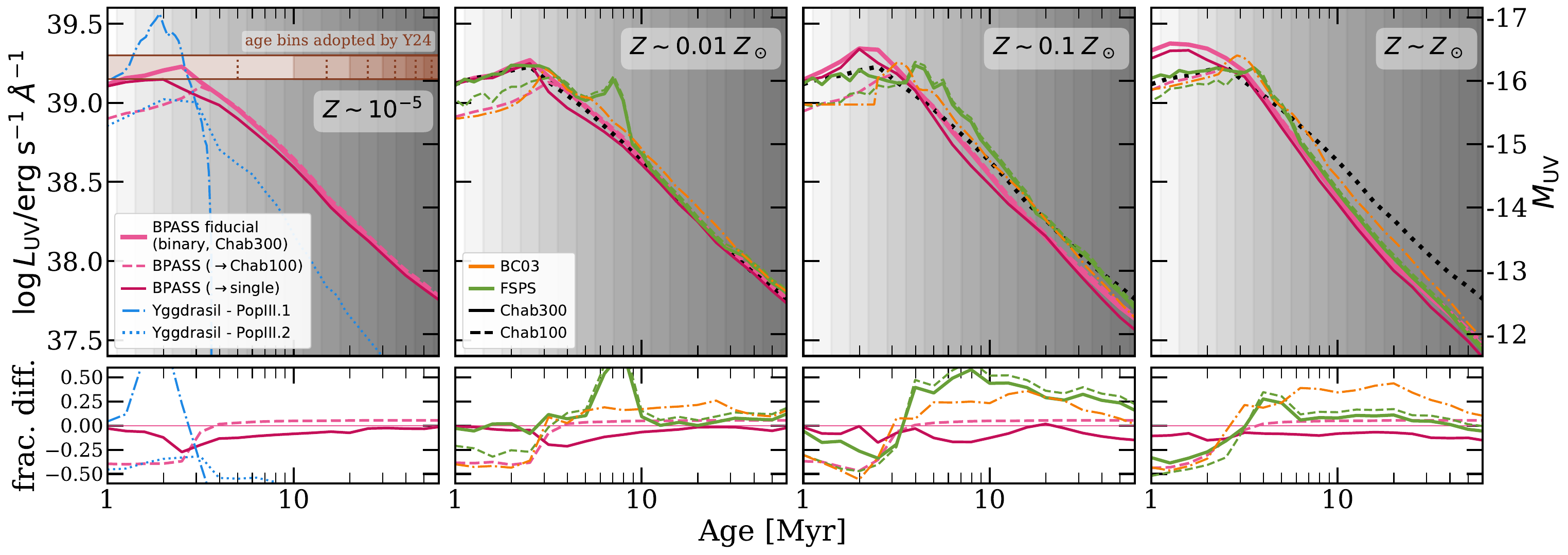}
    \caption{
        \textit{Top row:} The rest-frame FUV luminosity $L_\text{UV}$ (left axis) and magnitude (right axis) for a simple stellar population (SSP) of $M_* = 10^6$\Msun\ as a function of stellar age predicted by \textsc{bpass} (magenta and red for binary and single star models), \citetalias{Bruzual2003} (orange, Chabrier100 only), \textsc{fsps} (green), and the Yggdrasil (blue) spectral synthesis models, assuming instantaneous starbursts for metallicities of nearly metal-free ($Z \sim 10^{-5}$, \textit{left panel}), $\sim 0.01 Z_\odot$ \textit{(middle-left panel)}, $\sim 0.1 Z_\odot$ \textit{(middle-right panel)}, and solar metallicity $Z_\odot = 0.020$, \textit{right panel}). We indicate the IMF adopted by these models, with Chab300 and Chab100 indicating the use of a \citet{Chabrier2003} IMF with upper mass cut-offs at 300\Msun\ (solid lines) and 100 \Msun\ (dashed lines), respectively. We show the near-metal-free case of the \textsc{bpass} binary model with a Chabrier IMF with 300 \Msun\ cut-off in the other panels to guide the eye.
        The vertical grey shaded bars represent the boundaries of the re-binned SFHs adopted for constructing galaxy SEDs in this work, which centre around the stellar ages provided by \textsc{bpass}. We also mark the age bins adopted by \citetalias{Yung2024} with the bins and bin centres represented by the brown shaded regions and vertical dotted lines, respectively. This illustrates that the 10 Myr wide bins in linear stellar age are much wider than the 0.1 dex wide bins adopted by \textsc{bpass} and other SSP models, which reflect the physical timescales on which these stars evolve. \textit{Bottom row:} The fractional difference between $L_\text{UV}$ from various SSP models and the fiducial \textsc{bpass} configuration (binary, Chab300) adopted for the rest of this work. This figure highlights the significant differences between different SSP models at young ages, including the surprisingly large effect of adopting an upper mass cut-off of 300\Msun\ instead of 100 \Msun.
    }
    \label{fig:MUV_age}
\end{figure*}

\begin{figure*}
    \includegraphics[width=1.85\columnwidth]{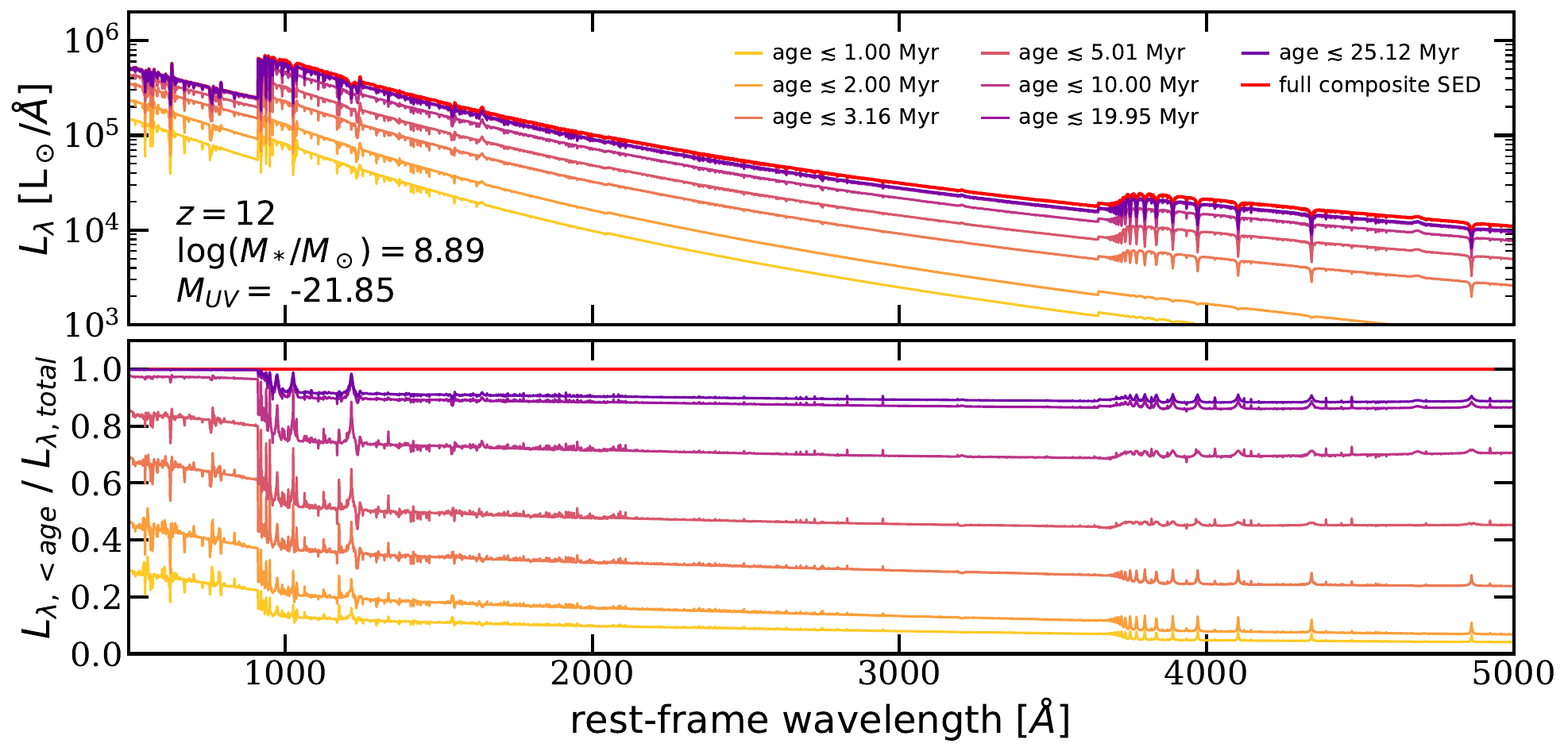}
    \caption{
        A breakdown of the composite stellar SED (\textit{top}) and the fraction of the total flux density (\textit{bottom}) that comes from stellar populations of different ages for a typical galaxy at $z\sim12$ with $\log(M_*/{\rm M}_\odot) = 8.89$ and a rest-frame $M_\text{UV} = -21.85$. We show that stars with age $<5$ ($<10$) Myr are responsible for $\sim 50$\% ($\sim 75$\%) of the total light emitted in the FUV.
    }
    \label{fig:SED_decomposition}
\end{figure*}

\subsection{Star formation histories and synthetic stellar spectra}
\label{sec:sfhist_sed}
 
The Santa Cruz SAM tracks star formation activity through the supplied merger histories of host halos and the physical processes summarized in Section~\ref{sec:sc-sam} and records the star formation rates in a two-dimensional grid of stellar age and metallicity. As noted in Section~\ref{sec:sc-sam}, when galaxies merge, their stellar populations are combined bin by bin. Rest-frame, unattenuated composite stellar spectra are then constructed by co-adding the spectra of single age, single metallicity ``simple stellar populations'' (SSPs), weighted appropriately, across these age and metallicity bins.

In this work and \citet{Yung2024}, we adopt stellar SEDs from the binary population and spectral synthesis \citep[\textsc{bpass}\footnote{\url{https://bpass.auckland.ac.nz/}, v2.2.1};][]{Stanway2016, Eldridge2017, Stanway2018}, which combine stellar isochrones with stellar atmosphere models or templates, weight them with an assumed stellar IMF, and provide stellar continuum SEDs for a grid of stellar ages and metallicities. The synthetic stellar SEDs assume a fiducial broken power law stellar IMF, with an upper slope $a_1 = -1.30$ between 0.1 -- 0.5 \Msun\ and a lower slope $a_2 = -2.35$ between $0.5$ and an upper mass cut-off of $m_\text{U} = 300$\Msun.
In Fig.~\ref{fig:MUV_age}, we illustrate the rest-frame far-UV luminosity (left y-axis) and magnitude (right y-axis) for a $M_* = 10^6$\Msun\ simple stellar population as a function of stellar age across a wide range of metallicities. Overall, the binary models yield brighter stellar populations than single star models, as expected. In Fig.~\ref{fig:MUV_age}, we also compare $M_{\rm UV}$--age relations from \textsc{bpass} to those from \citetalias{Bruzual2003} \citep{Bruzual2003}, Flexible Stellar Population Synthesis \citep[\textsc{fsps};][]{Conroy2010}, and the Yggdrasil Pop III models \citep{Zackrisson2011} where metallicity coverage overlaps. For young populations (ages of a few Myr), these models are in broad agreement in both normalization and qualitative behaviour.
It is evident that the integrated $M_\text{UV}$ can evolve very rapidly, approximately 4 magnitudes, over the first few tens of Myr. 
We also highlight that for very young stellar populations (e.g. stellar age $\lesssim 5$ Myr), the upper mass cut-off of the IMF has a significant impact on the integrated $M_\text{UV}$ of up to $\sim 0.5$ mag.
We note that several modelling assumptions can affect the predicted SEDs and $L_{\rm UV}$, including stellar rotation and rotational mixing, massive-star evolutionary tracks, and the adopted stellar-atmosphere and spectral libraries \citep{Meynet2000, Maeder2005, Conroy2009, Byrne2023}. These uncertainties are particularly difficult to calibrate in this regime because metal-poor massive stars are rare in the nearby Universe \citep{Telford2024}.

In previous work, \citet{Yung2024} constructed composite stellar spectra using age binning that matched the native internal binning choice of the Santa Cruz SAM, which records star formation in 10 Myr age bins. However, as illustrated in Fig.~\ref{fig:MUV_age}, the magnitude computed for the mean flux in the far-UV band from 1556\AA\ to 1576\AA\ is extremely sensitive to stellar age. For metal-poor populations ($Z \lesssim 10^{-4}$), $M_{\rm UV}$ can fade by $\sim$2 magnitudes between 1 and 20 Myr, and the evolution is even more rapid at higher metallicities.

In the left panel of Fig.~\ref{fig:MUV_age}, we illustrate the 10 Myr-wide bins adopted in \citet{Yung2024} (brown shaded bands), with vertical dotted lines marking the bin midpoints that are adopted when coupling with SSP SEDs. While this binning is adequate for galaxies in the $z \lesssim 10$ regime, whose SFHs typically extend over hundreds of Myr, it becomes problematic for ultra-high-redshift galaxies whose SFHs are dominated by very young stars. In this regime, the coarse age binning fails to capture the contribution from rapidly evolving young stellar populations, which dominate the rest-UV luminosity (see Section~3.2 and Figs.~\ref{fig:cumulative_mstar_z12} to \ref{fig:cumulative_mstar_allz_MUV}). As a result, \citet{Yung2024} systematically underestimated UV luminosities, with the severity increasing toward earlier times.

Fig.~\ref{fig:SED_decomposition} shows, for a sample galaxy with $\log(M_*/{\rm M}_\odot) = 8.89$ and a rest-frame $M_\text{UV} = -21.85$, drawn from \gureft-90 at $z\sim12$, the full composite stellar SED and a break down of the contribution by stellar populations of different ages. We show that stars with age $<5$ ($<10$) Myr are responsible for $\sim 50$\% ($\sim 75$\%) of the total light emitted in the FUV, which further motivates the need to properly account for the contributions from young stellar populations when computing UV luminosities.

To address this, we introduce an updated photometry computation routine that subdivides the SAM-output SFHs from the native 10 Myr bins into finer bins that are a factor of 15 narrower ($\simeq 0.67$ Myr). We tested a range of split factors and find that 15 provides a practical balance between temporal resolution and computational cost. The resulting fine SFH is then re-gridded onto the \textsc{bpass} age grid, which is evenly spaced in $\log$ age over $6 \leq \log({\rm age/yr}) \leq 11$ (vertical grey bands in Fig.~\ref{fig:MUV_age}). This procedure ensures that the contributions from the youngest stellar populations are properly accounted for in the composite stellar continuum spectra. Because the SAM does not include physical processes that drive star formation variability on timescales shorter than 10 Myr, storing SFHs at finer age resolution does not introduce new physical time variability beyond what is already present in the SAM; instead, it primarily improves the fidelity of the SFH-to-SED mapping. We discuss related caveats in Section~\ref{sec:discussion}.

Finally, we use the rest-frame composite spectra to compute rest-frame luminosities, accounting for dust attenuation in the interstellar medium (ISM) using a simple `slab' model, the same as the one adopted for the UV LF predictions by \citep{Yung2019}, that scales with metallicity and gas surface density \citep{Somerville2012}, with the redshift dependent normalization parameter calibrated as described by \citet{Yung2021}. Observed-frame magnitudes are then computed by redshifting the spectra and applying attenuation by the intervening IGM following \citet{Madau1996}.


\section{Results}
\label{sec:results}
In this section, we quantify the impact of the revised SFH-to-SED age-binning scheme (Section~\ref{sec:sfhist_sed}) on computed photometry and on the resulting rest-frame UV LFs. We then present an in-depth analysis of the predicted star formation histories of high- to ultra-high-redshift galaxies simulated with the Santa Cruz SAM coupled to merger trees from the \gureft\ suite.


\begin{figure*}
    \includegraphics[width=2\columnwidth]{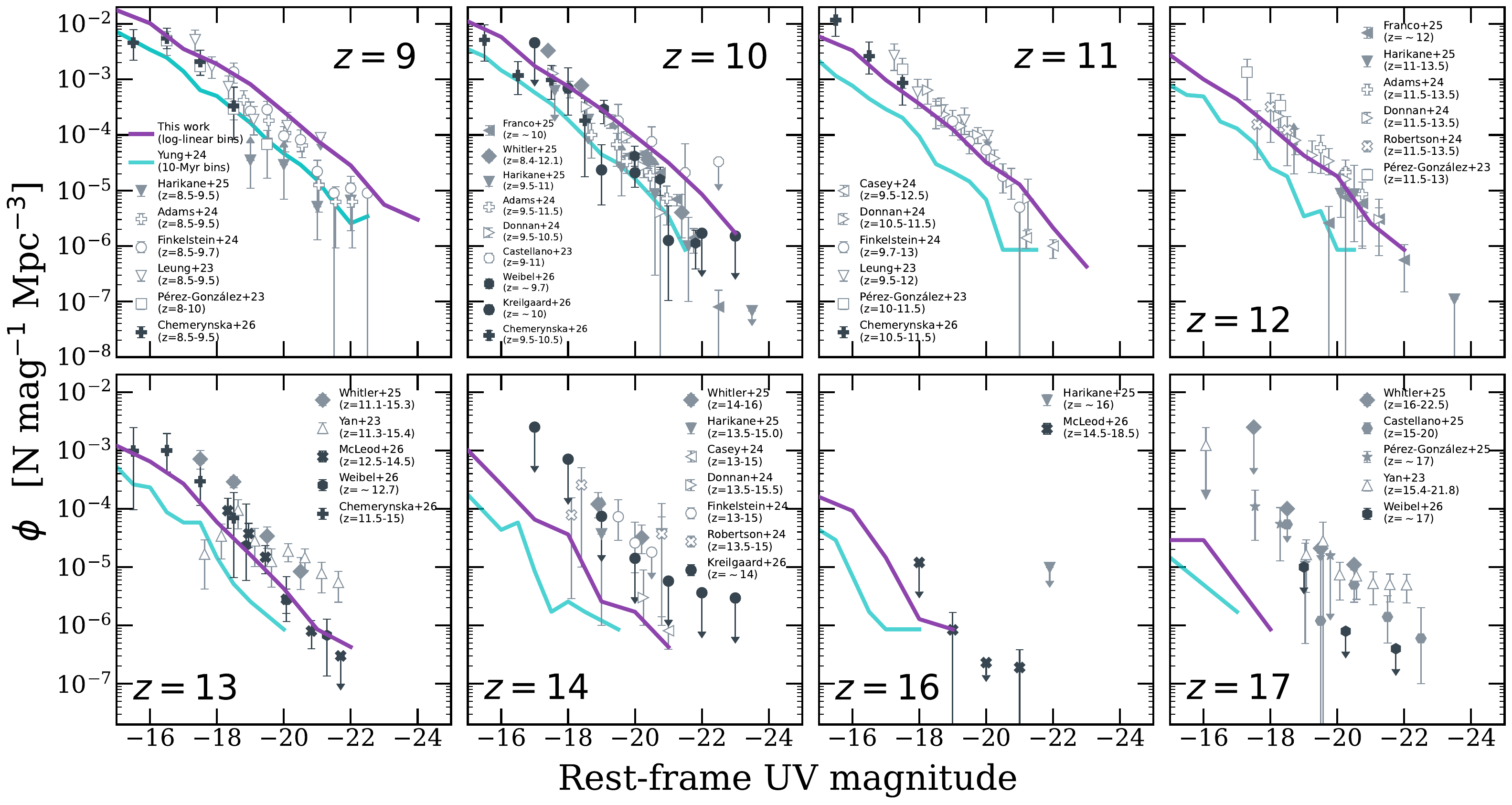}
    \caption{
        UVLFs at $z = 9$ to 17 run on \gureft\ merger trees, with photometry computed using 1 Myr-wide stellar age bins (solid purple lines), compared to the \citet{Yung2024} results, which utilized the 10 Myr-wide stellar age bins native to the Santa Cruz SAM outputs (cyan lines). 
        Both sets of UV LFs \textit{do not} include the effect of dust attenuation.
        The underlying simulations and predicted galaxy physical properties are identical between these runs. Grey symbols show a compilation of observational luminosity functions and the redshift range specific to these estimates, as specified in the figure legend (open symbols for results from 2024 and before, darker grey for recent results).
        \textit{Properly accounting for the contribution from young stellar populations is crucial for reproducing the observed UV luminous galaxy populations.}
    }
    \label{fig:UVLF_8panels_compY24}
\end{figure*}

\begin{figure*}
    \includegraphics[width=2\columnwidth]{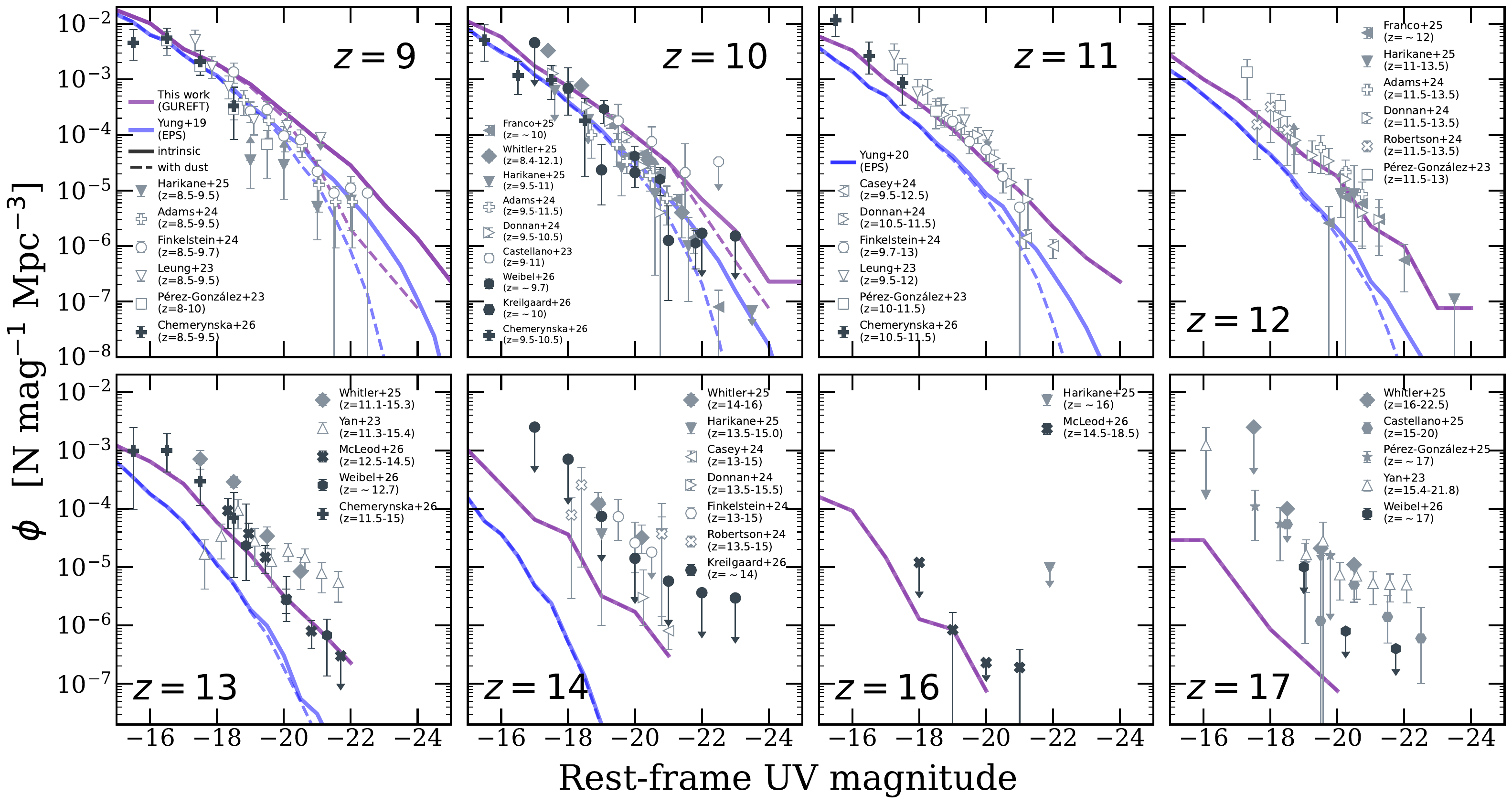}
    \caption{
        UVLFs at $z = 9$ to 17 with extension to more massive halos from VSMDPL without dust attenuation (solid purple line) and with dust attenuation assuming a `slab' dust model (dashed purple line). 
        Past simulations using EPS-based merger trees are shown for comparison \citep[blue solid and dashed lines for without and with dust attenuation, respectively;][]{Yung2019}.
        Grey symbols show a compilation of observational luminosity functions and the redshift range specific to these estimates, as specified in the figure legend (open symbols for results from 2024 and before, darker grey for recent results). \textit{Accurately capturing the underlying halo merger history is also critical for modelling ultra-high-redshift galaxies.}
    }
    \label{fig:UVLF_8panels_ext}
\end{figure*}

\subsection{Quantifying the impact on UV luminosity}
\label{sec:UVLFs}
Keeping the physical model and all free parameters fixed to those adopted in \citetalias{Yung2024}, we implement the updated age-binning scheme described in Section~\ref{sec:sfhist_sed}, which better captures the contribution from young stellar populations, particularly those with ages $<10$ Myr.
In Fig.~\ref{fig:UVLF_8panels_compY24}, we present a side-by-side comparison between UVLFs from \citetalias{Yung2024} and this work over $z=9$--17. Unless otherwise noted, these UVLFs are constructed from the intrinsic $M_{\rm UV}$ (without dust attenuation).
As in \citetalias{Yung2024}, the UVLFs are built by combining outputs across the four \gureft\ volumes. Here this procedure is automated following the steps and selection criteria described in Appendix~\ref{app:LF_combine_scheme}.
We compare to a compilation of recent \textit{JWST} constraints from deep extragalactic surveys \citep{Leung2023, Perez-Gonzalez2023, Yan2023, Adams2024, Casey2024, Donnan2024, Finkelstein2024, Robertson2024, Whitler2025, Perez-Gonzalez2025, Castellano2025, Franco2025, Harikane2025, Weibel2026, Kreilgaard2026, Chemerynska2026, Mcleod2026}.

This comparison shows that adopting $\sim$1 Myr-wide effective age resolution in the construction of composite spectra yields a $\sim$1 mag brightening in $M_{\rm UV}$ across the full redshift range considered, consistent with the age sensitivity of $M_{\rm UV}$ illustrated in Fig.~\ref{fig:MUV_age}. Importantly, this brings the predicted UVLFs into excellent agreement with the observed galaxy population up to $z \sim 12$ (and agrees with the lower UVLF estimates in the literature up to $z\sim 13$), despite adopting the identical underlying physical model as \citetalias{Yung2024}.
 
To extend the bright end, in Fig.~\ref{fig:UVLF_8panels_ext} we supplement the \gureft\ results with galaxies simulated in merger trees extracted from the larger Very Small MultiDark Planck simulation \citep[VSMDPL\footnote{\url{https://www.cosmosim.org/metadata/vsmdpl/}};][]{Klypin2016}. The inclusion of VSMDPL increases the sampled halo-mass range at the high-mass end and provides improved statistics for rare, UV-luminous systems relative to \gureft-90 alone. We also provide extended stellar mass functions, augmented by results from the larger VSMDPL volumes, in Appendix~\ref{app:extended_SMFs}. In Fig.~\ref{fig:UVLF_8panels_ext}, we show both intrinsic UVLFs and dust-attenuated UVLFs.
We note that this behaviour is consistent with the corresponding comparison in stellar mass functions shown in Fig.~2 of \citetalias{Yung2024}.

For context, Fig.~\ref{fig:UVLF_8panels_ext} compares our new results with results from the \textit{semi-analytic forecasts for JWST} series at $z\leq10$ \citep{Yung2019} and its extension to $z>10$ \citep{Yung2020a}, which used the same star formation prescriptions but were implemented within merger trees constructed using the extended Press--Schechter formalism \citep[EPS;][]{Press1974, Lacey1993, Somerville1999a, Somerville2008}. While the EPS approach is computationally efficient, the implementation relied on relatively sparse temporal sampling at $z>10$ and the results were never checked against numerical simulations at these redshifts. The discrepancy between EPS-based results and those based on \gureft\ is larger for the most luminous galaxies. This highlights that accurately capturing halo merger histories is a crucial prerequisite for modelling ultra-high-redshift galaxy populations.

Taken together, properly accounting for the contribution from young stellar populations (via refined SFH-to-SED age binning) and capturing the underlying halo merger history is sufficient to reproduce the observed UVLFs up to $z\sim 12$ within the current modelling framework. However, the same set of physical prescriptions is not sufficient to explain the observed galaxy populations at $z>13$. We refer the reader to \citet{Somerville2025} for an in-depth discussion and illustration of the role that additional physical processes, including density-modulated star formation efficiency, enhanced star formation stochasticity, and evolving dust attenuation, may play in accounting for the observed evolution of UV-luminous galaxies at these extreme redshifts. We note that the updated photometry method (with refined age bins) described here was already used in the work of \citet{Somerville2025}. 


\subsubsection{Relating SFR to rest-frame UV luminosity}

The conversion factor \KUV\ relates rest-frame UV luminosity to star formation rate, $\text{SFR} = {\cal{K}}_\text{UV} \times L_\nu(\text{UV})$. Many studies adopt the value of \KUV\ provided by \citet[][hereafter \citetalias{Madau2014}]{Madau2014}. 
The commonly adopted value of ${\cal{K}}_\text{UV}=1.15 \times 10^{-28}$ \Msun\ yr$^{-1}$\,/\,erg s$^{-1}$ Hz$^{-1}$ was computed by assuming an empirical, parametrized form for galaxy SFH and chemical enrichment history, with the onset of star formation activity at $z=12$ and results for \KUV\ presented for $z=0$--8 (see Fig.~3 of \citetalias{Madau2014} and the accompanying discussion, and Appendix~\ref{app:KUV_MD14} of this work). This calculation furthermore assumed a \citet{Salpeter1955} IMF over $0.1$--100\,\Msun\ and adopted the flexible stellar population synthesis (FSPS) models of \citet{Conroy2009}. Similar results were obtained for assumptions of a constant star formation rate and constant metallicity for ages of $\gtrsim 300$ Myr (${\cal{K}}_\text{UV}= $ (1.55, 1.3, 1.1, 1.0) $\times 10^{-28}$ \Msun\ yr$^{-1}$\,/\,erg s$^{-1}$ Hz$^{-1}$ for $\log (Z_{*}/Z_{\odot}) =$ (+0.2, 0, $-0.5$, $-1.0$; \citetalias{Madau2014}).  We further discuss the methodology adopted by \citetalias{Madau2014} and how their computation compares with the one adopted in this paper in Appendix \ref{app:KUV_MD14}.

In \citet{Yung2024}, \KUV\ was computed using the SAM-predicted SFR averaged over 100~Myr and $L_\nu(\text{UV})$ was obtained by integrating the composite stellar SEDs. However, those results suggested that galaxies were fainter (at fixed SFR) than implied by the \citetalias{Madau2014} conversion, which is the opposite of the expectation that the younger, more metal-poor stellar populations typical of the ultra-$z$ Universe should yield \emph{higher} UV light-to-mass ratios. 
 
\begin{figure}
	\centering
	\includegraphics[width=0.97\columnwidth]{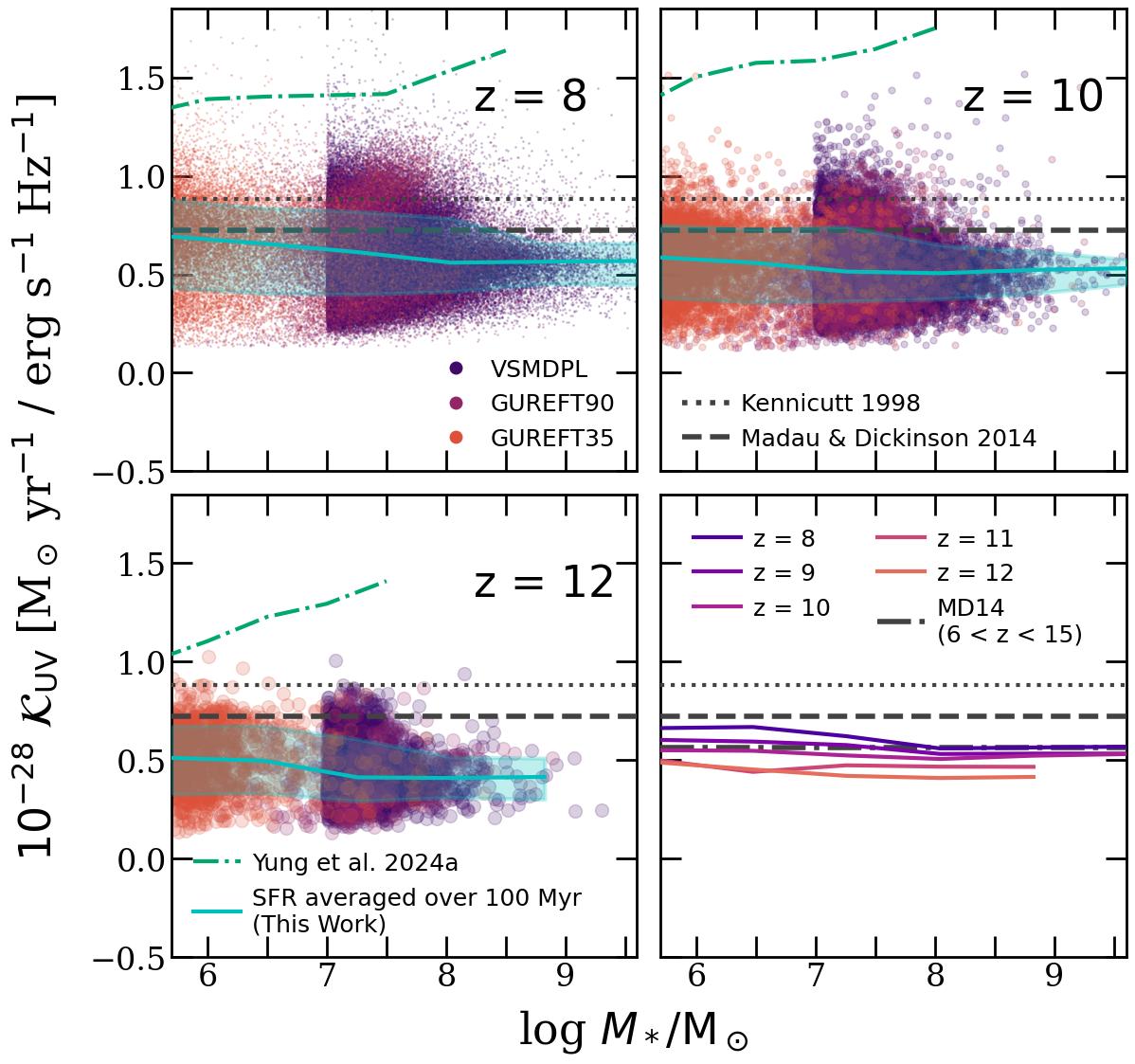}
	\caption{
		The UV-to-SFR conversion factor, \KUV, as a function of stellar mass for simulated galaxies from \gureft-90 and \gureft-35 at $z=8$, 10, and 12. We compute \KUV\ using $\text{SFR}_{100}$, matching the definition adopted in \citet{Yung2024} to enable a direct comparison. Because the updated photometry pipeline yields brighter rest-frame UV luminosities at fixed SFR, the inferred \KUV\ values are systematically lower than those reported in \citet{Yung2024}, shown by the green dot-dashed lines. The lower-right panel shows the redshift evolution of the median relation over $8 < z < 12$. For comparison, we also show the \KUV\ value computed using the star formation and metallicity histories adopted by \citetalias{Madau2014} together with \textsc{bpass} SEDs assuming a Chabrier IMF and an upper stellar mass cut-off of 300~\Msun, consistent with the assumptions adopted in this work, and evaluated over the redshift range $6 < z < 15$ (long dot-dashed line). See Appendix~\ref{app:KUV_MD14} for details.
	}
	\label{fig:KUV_mstar}
\end{figure}

In Fig.~\ref{fig:KUV_mstar}, we recompute the \KUV--$M_*$ relation presented in \citet{Yung2024}, using the same approach, but replacing the photometry with the updated pipeline described in Section~\ref{sec:UVLFs}.
This demonstrates that, for the same SFR, the revised SFH-to-$L_{\text{UV}}$ mapping yields more UV light, and therefore lower inferred \KUV. For a direct, like-for-like comparison to \citet{Yung2024}, we continue to compute \KUV\ using $\text{SFR}_{100}$ (SFR averaged over 100 Myr). However, as we discuss in Section \ref{sec:smah}, many galaxies at $z\sim 12$ form the bulk of their stars on timescales significantly shorter than 100 Myr, making $\text{SFR}_{100}$ a less physically meaningful descriptor of their star formation activity. We note that the value of \KUV\ shown as the \citetalias{Madau2014} value in Fig.~\ref{fig:KUV_mstar} is adjusted from a Salpeter to Chabrier IMF by multiplying by a factor of 0.63, as prescribed by \citetalias{Madau2014}.

\subsection{Ultra-high-redshift star formation histories and star formation efficiencies}
\label{sec:results:SFE}
\begin{figure*}
    \includegraphics[width=1.8\columnwidth]{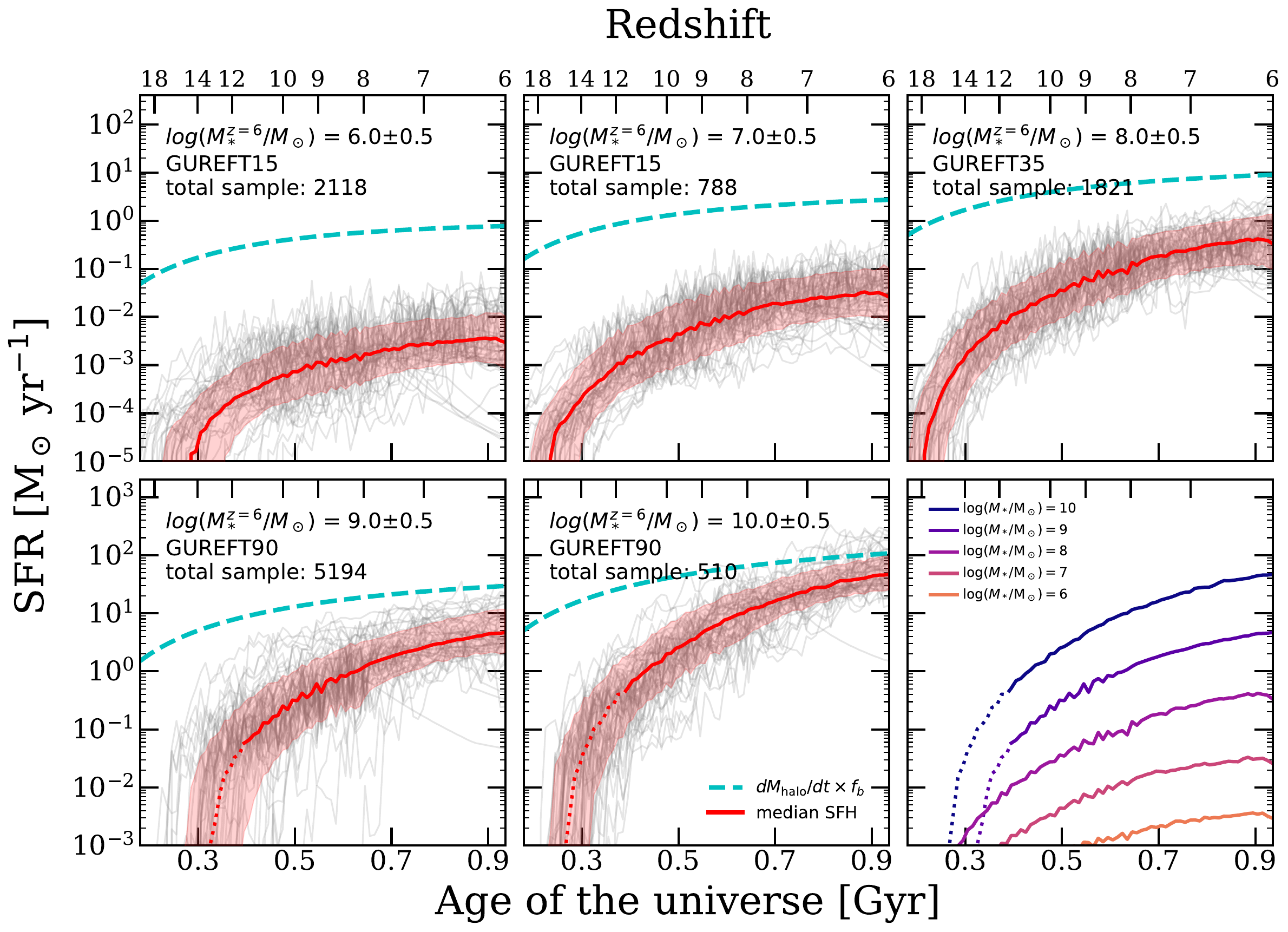}
    \caption{
        Star formation histories (SFHs) as a function of the age of the Universe (bottom $x$-axis) or redshift (top $x$-axis) for galaxies predicted by the Santa Cruz SAM, binned by their terminal stellar masses at $z=6$. We show the SFHs of a random sample of 50 SFHs in grey to illustrate the diverse SFH of individual galaxies. Red solid lines and shaded regions indicate the median and the 16th to 84th percentile range within each mass bin, respectively, computed for all available galaxies. In these panels, we overplot the halo mass accretion histories (MAHs) corresponding to the median host halo mass for galaxies in each terminal stellar mass bin, computed using the analytic fitting function from \citet{Yung2024a}. The final panel overlays the median stellar mass accretion histories across all mass bins for comparison. We note that halos sampled from the \gureft-90 simulation may have prematurely truncated merger trees due to limited mass resolution, which would cause the predicted SFR to be underestimated, and are represented with dotted lines. This is further discussed in Section \ref{sed:caveats_dynamicrange}.
    }
    \label{fig:SFH_mstar}
\end{figure*}

In this sub-section, we investigate the full star formation histories (SFHs) of galaxies in the high-$z$ to ultra-high-$z$ regime and explore the implied evolution of star formation efficiency (SFE) over a galaxy's growth history. We identify galaxies based on their global properties (such as stellar mass or luminosity) at the redshift where they are output and stored, which we refer to as the \emph{terminal redshift}. We emphasize again that throughout this work, SFH are constructed by plotting the summed SFR or stellar mass of \emph{all of the progenitor galaxies that end up in the descendent} galaxy at the specified terminal redshift. In this sense, the SFH include contributions from both \textit{in-situ} star formation (formed in the main progenitor) and \textit{ex-situ} star formation (formed in progenitors that later merge into the main system). Thus, these SFH are directly comparable to those that can be extracted from observed SEDs. 

In Fig.~\ref{fig:SFH_mstar} we present SFHs as a function of cosmic time or redshift for our SAM galaxies, binned by the galaxy stellar masses at $z=6$. This representation provides a view of how the star formation rate evolves within populations that reach comparable stellar masses by the same terminal epoch. Galaxies in different terminal-mass bins are drawn from different \gureft\ volumes, as annotated in each panel of Fig.~\ref{fig:SFH_mstar}. Our selection balances the competing requirements of mass resolution (which favours smaller boxes) and sample size (which favours larger boxes). 

As the terminal mass increases across panels, we sample halos from different \gureft\ boxes in order to balance merger-tree mass resolution against statistical robustness \citep[cf.][]{Yung2024a}. For instance, in overlapping halo-mass ranges, smaller boxes typically provide better-resolved merger histories, while larger boxes provide larger samples (see Fig.~\ref{fig:cSFH_mhalo_G35G90_app}). For the results in Fig.~\ref{fig:cumulative_mstar_z6}, we require that the terminal halo contains at least 120 DM particles (see Fig.~\ref{fig:cSFH_mhalo_z12_app} and discussion in Appendix \ref{app:resolution_test}), which we find that the minimum resolution needed to yield smooth recent-time behaviour in the predicted SFHs. For each panel, we select the simulation volume that satisfies this resolution requirement while maximizing the sample size, and we annotate the adopted \gureft\ box and the number of galaxies available for reference.

We note that we do not expect modest differences in halo mass resolution across \gureft\ volumes to strongly affect the predicted galaxy populations in mass ranges where the simulations overlap, as both the halo populations and the SAM-predicted galaxy statistics converge across boxes \citep{Yung2024, Yung2024a}, and additional tests of halo assembly statistics have been presented for \gureft\ merger trees \citep{Nguyen2024, Nguyen2025}. However, mass resolution becomes more important for the specific questions addressed here, particularly the onset of star formation at very early times, because resolving early progenitor halos directly determines whether the earliest star formation episodes are captured.

In Fig.~\ref{fig:SFH_mstar}, SFHs of individual galaxies are shown in grey, illustrating the galaxy-to-galaxy stochasticity in SFH due to merger-triggered bursts, past mergers of objects with distinct stellar populations, and the baryon feedback cycle, as discussed in Section~\ref{sec:sc-sam}. Although individual SFHs show great diversity, the median and the 16th to 84th percentile range (red solid lines and shaded regions) reveal systematic trends with terminal stellar mass. We find that SFR of galaxies generally increase as a function of time, and the overall growth of SFR over time strongly correlates with the terminal stellar mass, where SFR grows more rapidly for more massive galaxies. This is expected as it has been shown that SFR and $M_*$ are strongly correlated \citep[e.g.][and references therein]{Speagle2014, Popesso2023}. 
For the most massive systems, which are only present in \gureft-90, resolving the earliest onset of star formation is somewhat limited by the coarser mass resolution of the underlying merger trees. However, because star formation ramps up rapidly once these massive halos are established, any unresolved early-time activity is expected to have limited impact on the final stellar masses and bulk galaxy properties. The impact of mass resolution on cumulative stellar mass growth is further explored in Appendix \ref{app:resolution_test}.

To place these SFHs in the context of halo growth, in Fig.~\ref{fig:SFH_mstar} we overlay halo mass accretion histories (MAHs) corresponding to the median host halo mass for galaxies in each terminal stellar mass bin. We approximate the \emph{average} halo growth using the fitting function from \citet{Yung2025}:
\begin{equation}
    dM_\text{h}/dt(M_\text{h},z) = \beta(z)(M_\text{h,12}E(z))^{\alpha(z)}\text{,}
\end{equation}
where $M_{{\rm h},12} \equiv M_{\rm h}/(10^{12}\,{\text{M}_\odot})$. Based on halo accretion rates measured in \gureft, the redshift-dependent parameters $\alpha(z)$ and $\beta(z)$ are well described by
\begin{equation}
\begin{split}
    \alpha(z)    &= 0.948 + 0.694 a - 0.565 a^2\\
    \log\beta(z) &= 2.673 - 2.075 a + 0.891 a^2 \text{,}
\end{split}
\end{equation}
where $a \equiv 1/(1+z)$ is the scale factor and $E(z)$ is the dimensionless Hubble parameter.

\begin{figure*}
    \includegraphics[width=1.25\columnwidth]{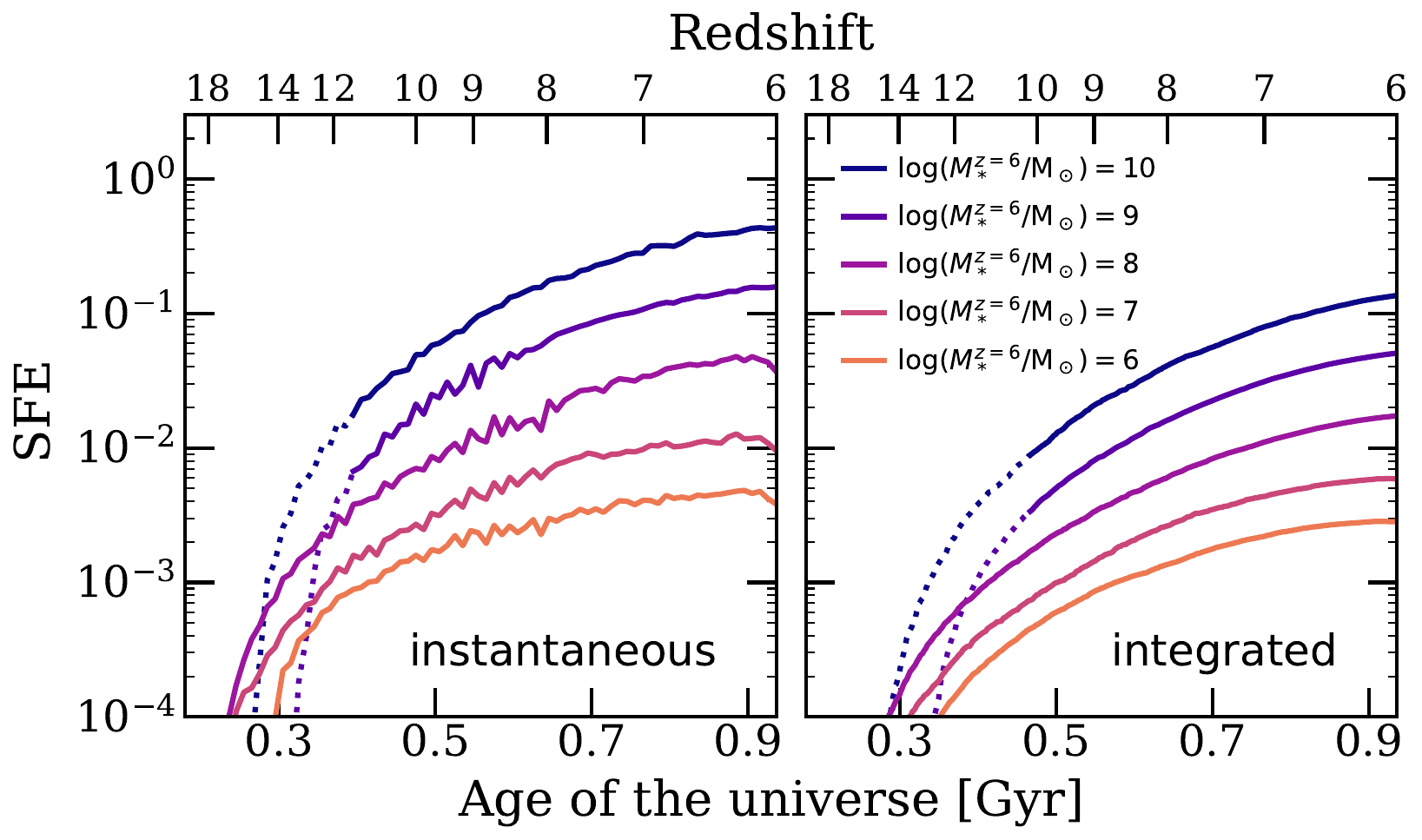}
    \caption{
        Median instantaneous (\textit{left}) and integrated (\textit{right}) star formation efficiencies (SFEs) as a function of the age of the universe (bottom $x$-axis) or redshift (top $x$-axis)  for galaxies binned by their terminal stellar masses at $z=6$. We find that more massive galaxies identified near the end of the EoR maintain higher star formation efficiencies throughout their histories. 
    }
    \label{fig:SFE_evolution}
\end{figure*}

Using the median SFHs in Fig.~\ref{fig:SFH_mstar} together with the MAH approximation above, we examine how SFE evolves over the course of galaxy growth. In Fig.~\ref{fig:SFE_evolution}, we show two commonly used definitions: 1) the \textit{instantaneous} SFE, $\text{SFE}_\text{inst} \equiv \dot{m}/(f_b\dot{M_\text{h}})$, where $\dot{m}$ is the SFR, $f_b$ is the universal baryon fraction, and $\dot{M_\text{h}}$ is the halo mass accretion rate; and 2) the \textit{integrated} SFE, $\text{SFE}_\text{int} \equiv m_*/(f_bM_h)$, where $m_*$ is the stellar mass and $M_\text{h}$ is the host halo mass.

While it is expected that more massive galaxies (and halos) tend to have higher SFEs \citep[see also Fig.~9 in][as well as observational constraints therein]{Somerville2025}, we find that, for galaxies with similar terminal masses, both their median $\text{SFE}_{\rm inst}$ and $\text{SFE}_{\rm int}$ generally increase with time along their evolutionary tracks. This implies that more massive galaxies identified at a given epoch \emph{have had higher SFE throughout the course of their evolution} compared to their lower-mass counterparts. This is due to the higher assumed mass loadings for stellar driven winds in lower mass halos in the Santa Cruz SAM, as discussed further in Section~\ref{sec:SFE_evolution}.

\begin{figure}
    \includegraphics[width=1\columnwidth]{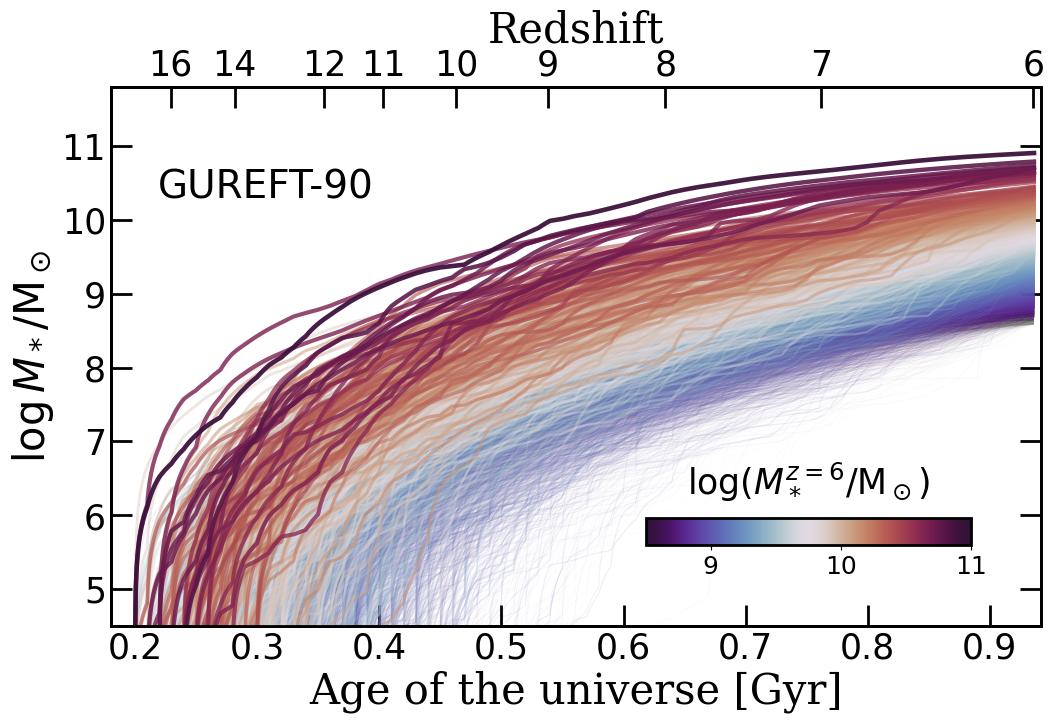}
    \caption{
        Overview of stellar mass assembly histories (SMAHs) for galaxies terminating at $z=6$ with $8.5 \lesssim \log(M^{z=6}_*/{\rm M_\odot}) \lesssim 11$, simulated with the Santa Cruz SAM in merger trees extracted from \gureft-90. Each track shows the cumulative stellar mass as a function of cosmic time (top axis) and redshift (bottom axis), and is colour-coded by the terminal stellar mass at $z=6$. This figure is designed to illustrate the \textit{decorrelation} between terminal masses and progenitor growth tracks: galaxies with similar $M^{z=6}_*$ can have widely different assembly histories. For visual clarity, we plot thicker, more opaque lines for rarer, more massive systems and thinner, more transparent lines for more abundant, lower-mass systems. The increasing degree of track crossing and mixing toward higher redshift highlights the weakening correspondence between terminal mass and progenitor mass at early times.
    }
    \label{fig:smah_overview}
\end{figure}

\begin{figure*}
	\includegraphics[width=1.8\columnwidth]{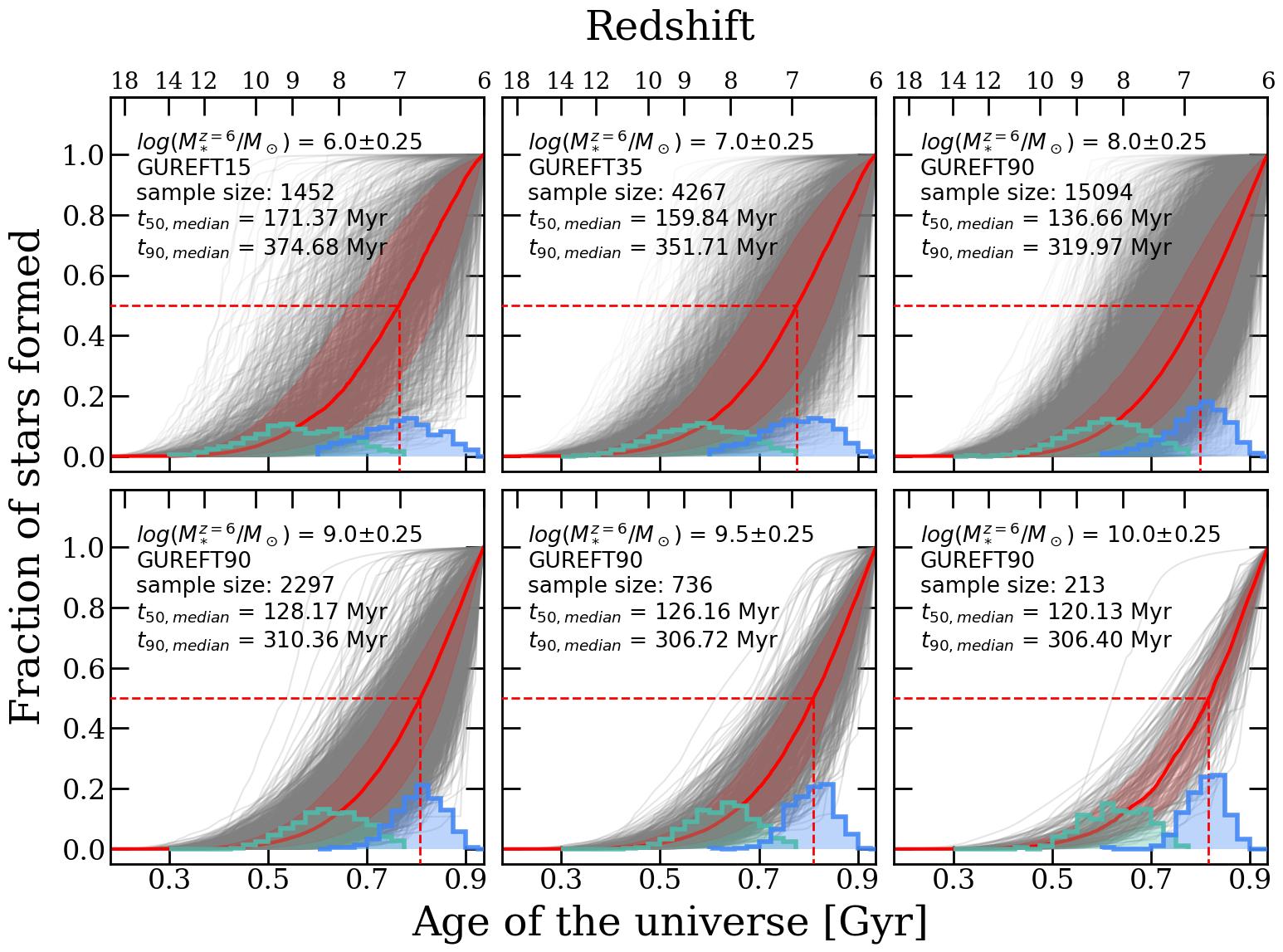}
	\caption{
		Fraction of stars formed for galaxies in various stellar mass bins, normalized to their final stellar mass at $z=6$, as a function of cosmic time (bottom axis) and redshift (top axis), simulated with the Santa Cruz SAM in various \gureft\ volumes (see annotated text in each panel). Individual galaxies are shown in grey. The red solid line and shaded region represent the median and the 16th to 84th percentile range, respectively, illustrating the spread across the population. The vertical dashed lines indicate the redshifts at which the median cumulative stellar mass reaches 50 percent of its final stellar mass. The cyan and blue histograms in each panel show the distribution of $t_{90}$ and $t_{50}$, which are the amount of time required for galaxies to form the final 50 per cent and 90 per cent of their stellar mass, respectively. 
	}
	\label{fig:cumulative_mstar_z6}
\end{figure*}

\begin{figure*}
	\includegraphics[width=1.8\columnwidth]{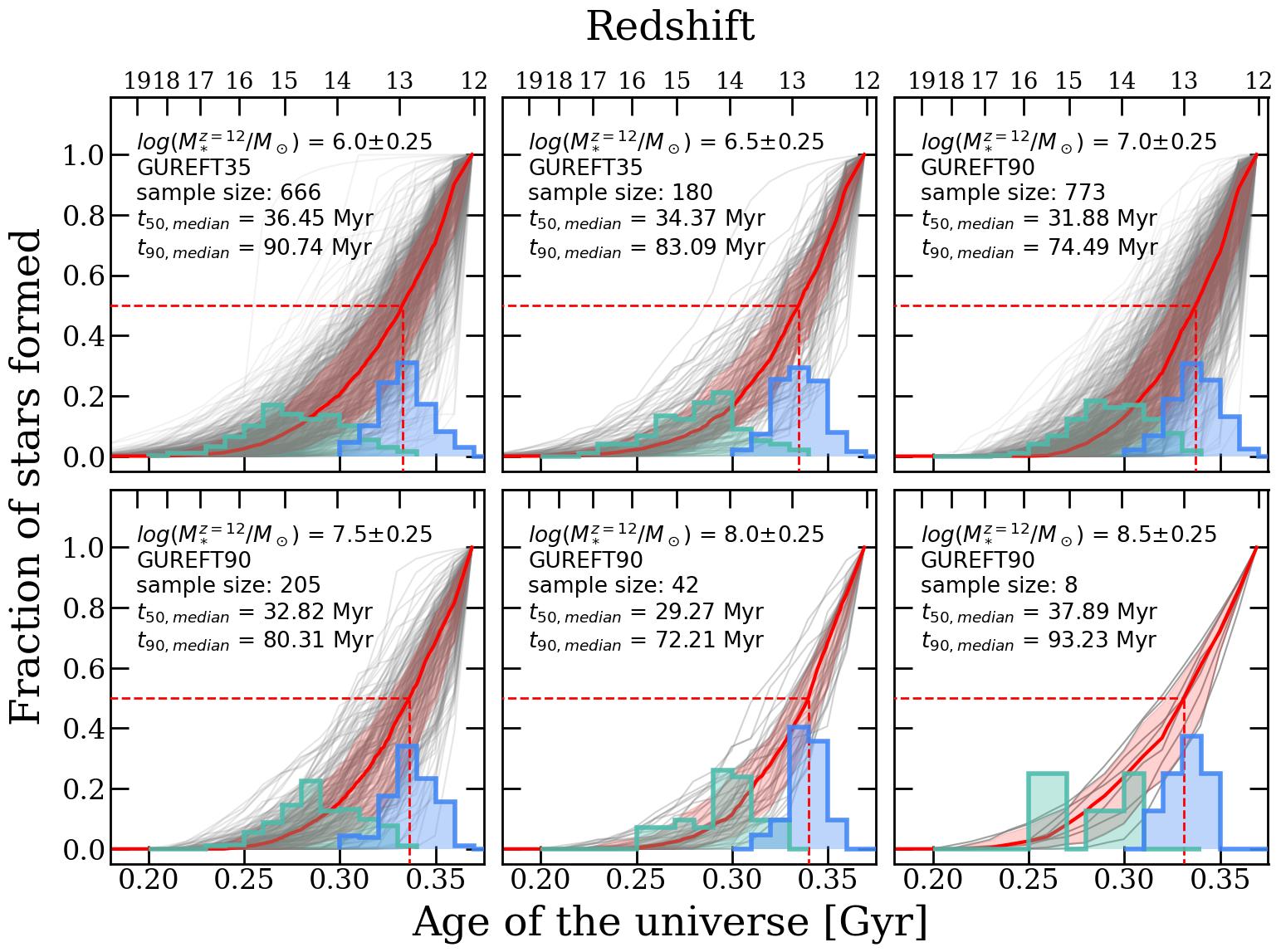}
	\caption{
		Fraction of stars formed for galaxies in various stellar mass bins, normalized to their final stellar mass at $z=12$, as a function of cosmic time (bottom axis) and redshift (top axis), simulated with the Santa Cruz SAM in various \gureft\ volumes (see annotated text in each panel). Individual galaxies in the sample are shown in grey. The red solid line and shaded region represent the median and the 16th to 84th percentile range, respectively, illustrating the spread across the population. The vertical dashed lines indicate the redshifts at which the median cumulative stellar mass reaches 50 percent of its final stellar mass. The cyan and blue histograms in each panel show the distribution of $t_{90}$ and $t_{50}$, which are the amount of time required for galaxies to form the final 50 per cent and 90 per cent of their stellar mass, respectively.
	}
	\label{fig:cumulative_mstar_z12}
\end{figure*}

\subsection{Stellar age distributions and star formation timescales}
\label{sec:smah}

In Fig.~\ref{fig:smah_overview}, we provide an overview of cumulative stellar mass growth as a function of cosmic time and redshift for galaxies simulated in merger trees extracted from the \gureft-90 volume, terminating at $z\simeq 6$. We colour-code each assembly track by its terminal stellar mass, $M^{z=6}_*$. While the tracks are loosely correlated with terminal mass, they exhibit substantial diversity even among galaxies that reach similar $M^{z=6}_*$, illustrating that terminal mass alone does not uniquely determine a galaxy's progenitor growth track.

In Fig.~\ref{fig:cumulative_mstar_z6}, we break down the sample galaxy population by their terminal stellar mass at $z\simeq6$, $M^{z=6}_*$, and bin together galaxies of similar masses, using bins of width $\Delta \log(M^{z=6}_*/{\rm M_\odot}) = 0.5$. For ease of comparison across mass bins, we present the \emph{fraction of stellar mass formed} as a function of time since the Big Bang (each SFH normalized by the galaxy's terminal mass). As discussed in the previous section, we sample halos from different \gureft\ boxes for different terminal masses. 

In each panel of Fig.~\ref{fig:cumulative_mstar_z6}, individual normalized SFH histories are shown in grey. We characterize the scatter among among these populations by marking the median and the 16th to 84th percentile range with the red solid line and shaded region. We also mark, with a red dashed line, the time at which the \emph{median} stellar mass formed reaches 50 per cent of the terminal stellar mass.

To quantify formation timescales in a compact way, we define $t_X$ as the lookback time (relative to the terminal redshift) over which a galaxy formed its most recent $X$ per cent of stellar mass. For instance, $t_{50}$ is the lookback time over which the galaxy formed the most recent 50 per cent of its terminal mass, and $t_{90}$ is the lookback time over which it formed the most recent 90 per cent. These quantities are equivalent to the median (or 10th or 90th percentiles) of the \emph{mass-weighted ages} of the stellar populations in these galaxies. We use $t_{50}$ as a characteristic timescales for recent growth (and therefore sensitivity to young stellar populations), and $t_{90}$ as a proxy for the timescales associated with forming the bulk of the stellar mass. We note that the convention adopted in this work is different from some past studies \citep[e.g.][]{Pacifici2016}, which is deliberately chosen to anchor these timescales to the terminal mass at the epoch of observation. In Fig.~\ref{fig:cumulative_mstar_z6}, we show the distributions of $t_{50}$ and $t_{90}$ over the full population in the plotted terminal mass bin as histograms, and annotate their medians in each panel. We find that galaxies terminating at $z\simeq 6$ have $t_{50} \sim 120$--170~Myr, with more massive systems having shorter values of $t_{50}$ on average. We also note that $t_{50}$ computed from the median SFH provides a good approximation to the median of the galaxy-by-galaxy $t_{50}$ distribution. 

In Fig.~\ref{fig:cumulative_mstar_z12}, we repeat the same analysis for galaxies identified and binned by their stellar mass at $z\simeq12$, motivated by the ultra-high-redshift galaxy populations recently detected with \textit{JWST} \citep[e.g.][]{Finkelstein2022, Finkelstein2023, Finkelstein2024, Leung2023a}. The resulting SFH are noticeably more compressed in time. Across the mass range shown, most simulated ultra-$z$ galaxies form $\sim$50 per cent of their stellar mass within the past $\sim 35$~Myr, which is only $\sim$20 per cent of the time required for $z\simeq 6$ galaxies of comparable stellar mass to form the same fraction. This highlights that star formation timescales evolve rapidly with redshift, and that the characteristic growth timescales of ultra-$z$ galaxies were substantially shorter than those of their counterparts near the end of the EoR. This compression of formation timescales also underscores the importance of properly modelling very young stellar populations, both in galaxy scale simulations and in stellar population models. 

In Appendix~\ref{app:SMAH_fitting_function}, we provide a functional form that provides a good description of the cumulative star formation histories in the SC SAM over the redshift range $6 \lesssim z \lesssim 10$. As shown there, we find that a bounded power law form provides a better fit to the data than an exponential, particularly toward the lower-redshift end of this interval. This reflects the trends in SFH shown in Fig.~\ref{fig:SFH_mstar}, where star formation rates rise more rapidly at early times and gradually slows down towards lower redshift.

We note that, as shown in Figs.~\ref{fig:SFH_mstar}, \ref{fig:cumulative_mstar_z6}, and \ref{fig:cumulative_mstar_z12}, galaxies with similar terminal stellar masses can have very diverse  evolution histories. The median and the 16th to 84th percentiles provide a useful summary of the overall evolutionary trend across the full galaxy population, but may not be a good representation of the evolution history of individual galaxies (in particular, they are inherently more smooth due to being averaged over many objects).


\begin{figure}
    \includegraphics[width=0.9\columnwidth]{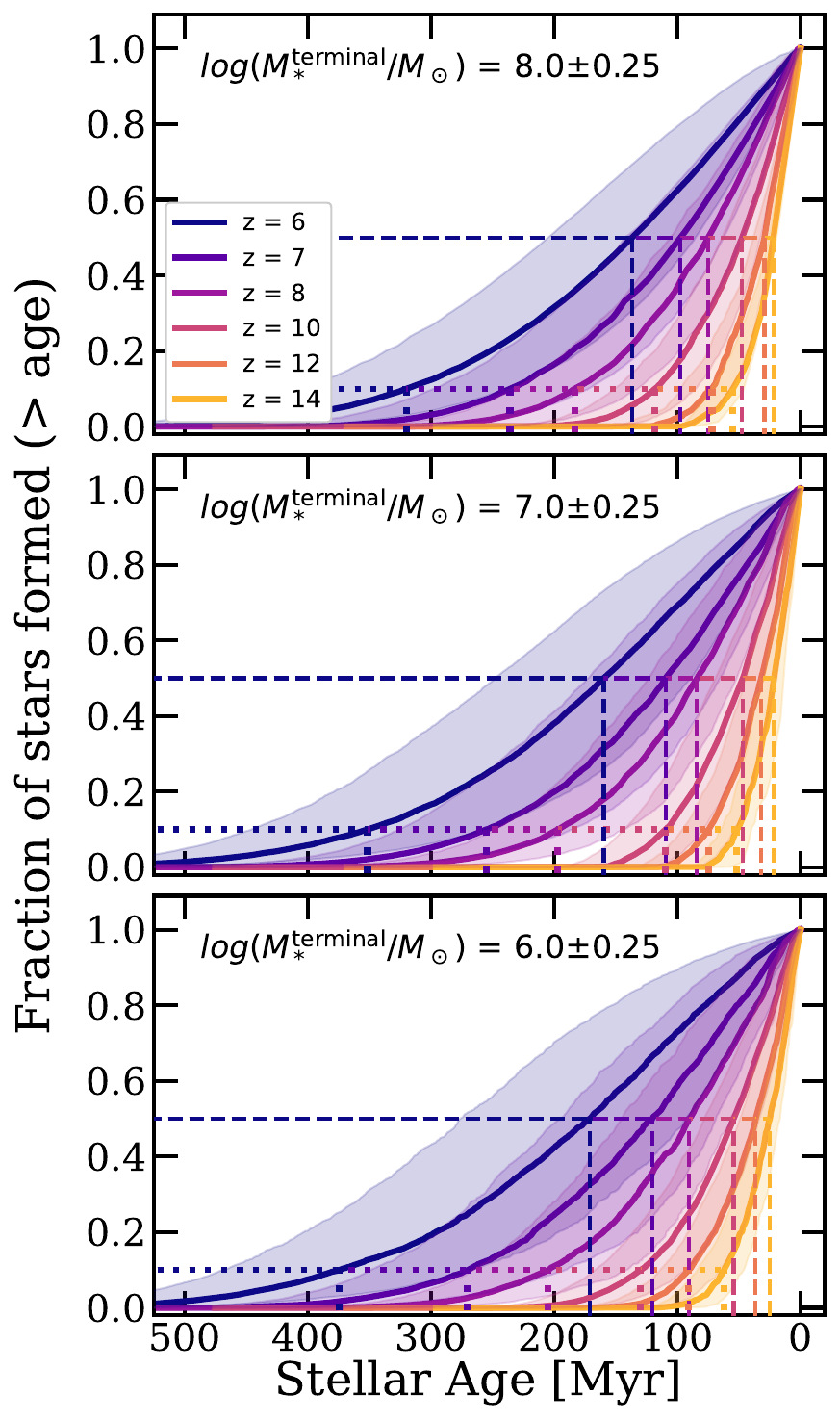}
    \caption{
        Normalized stellar mass assembly histories (cumulative fraction of stellar mass formed) for galaxies identified at $z = 6$, 7, 8, 10, 12, and 14 (see legend for colour coding) in bins of terminal stellar mass with $\Delta \log(M_*/\text{\Msun}) = 0.5$ as annotated in each panel, shown as a function of stellar age as predicted by the Santa Cruz SAM. Solid lines and shaded regions indicate the median and the 16th to 84th percentile range, respectively. Vertical dashed and dotted lines mark the lookback times corresponding to $t_{50}$ and $t_{90}$, respectively. This figure illustrates that galaxies with similar terminal stellar masses can exhibit substantially different assembly histories, with systematic differences across cosmic epochs, particularly at early times.
    }
    \label{fig:cumulative_mstar_allz}
\end{figure}

\begin{figure}
    \includegraphics[width=0.9\columnwidth]{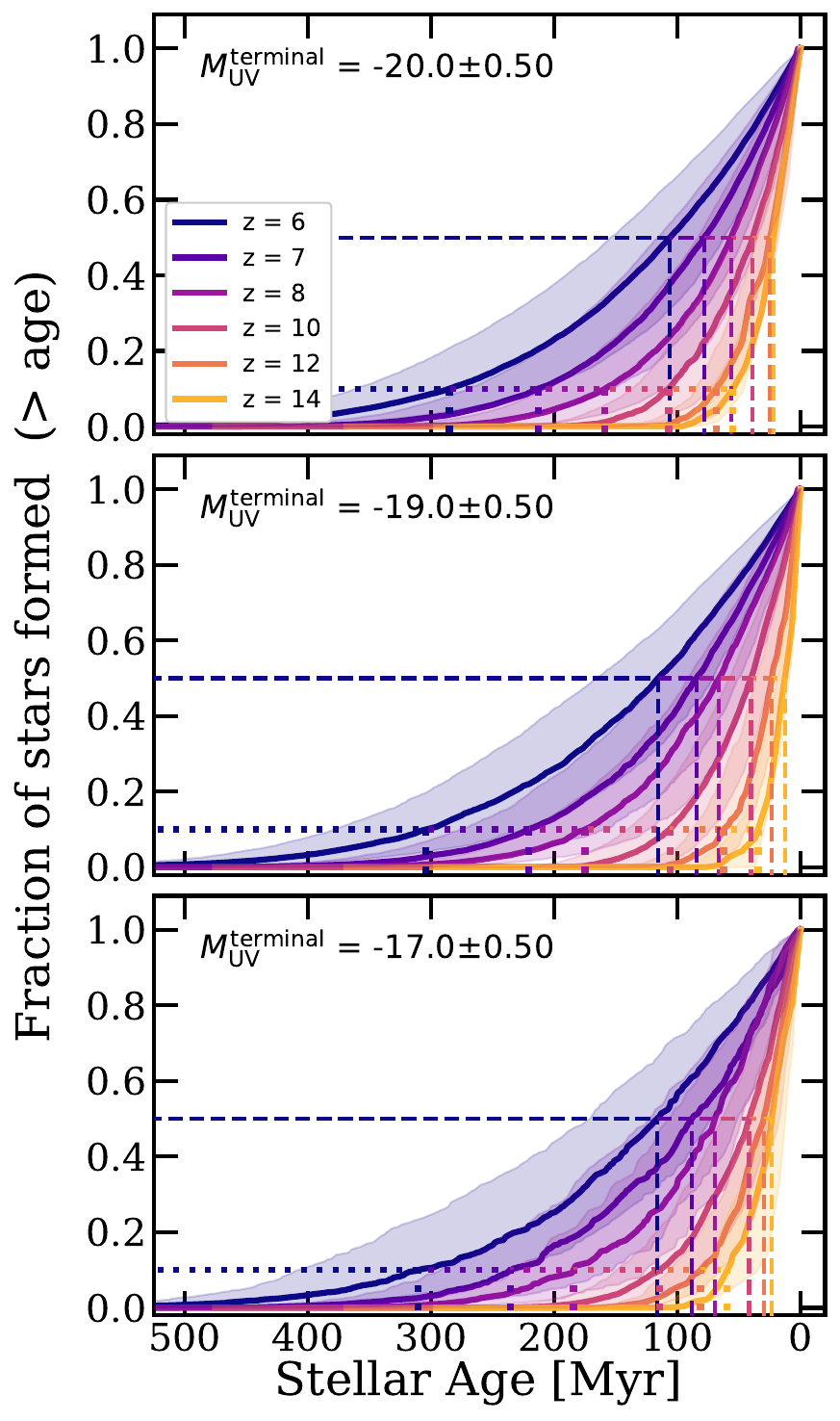}
    \caption{
        Normalized stellar mass assembly histories (cumulative fraction of stellar mass formed) for galaxies at $z = 6$, 7, 8, 10, 12, and 14 (see legend for colour code) in bins of rest-frame, dust attenuated UV luminosity bins at the epoch of observation with $\Delta M_\text{UV} = 1.0$, shown as a function of stellar age as predicted by the Santa Cruz SAM. The solid lines and shaded regions mark the median and 16th and 84th percentile range. Vertical dashed and dotted lines mark the lookback times corresponding to $t_{50}$ and $t_{90}$, respectively. This figure highlights that galaxies with similar $M_{\rm UV}$ at different redshifts can exhibit substantially different assembly histories, with systematic differences across cosmic epochs, particularly at early times. However, the characteristic assembly timescales do not exhibit a strong dependence on $M_{\rm UV}$.
    }
    \label{fig:cumulative_mstar_allz_MUV}
\end{figure}

In Fig.~\ref{fig:cumulative_mstar_allz}, we further explore these trends by compiling star formation histories for galaxies normalized to their terminal stellar mass (equivalent to fraction of stellar mass formed) over terminal redshifts $6 \lesssim z \lesssim 14$ in bins of terminal stellar mass with $\Delta \log(M_\ast/{\rm M_\odot})=0.5$, and present them as a function of lookback time relative to their epoch of observation, such that all curves are aligned at $t_{\rm lookback}=0$. As in the previous figures, we indicate $t_{50}$ and $t_{90}$ for the \emph{median} SFH  in each redshift slice, marking them with dashed and dotted vertical lines coloured by redshift. 

In Fig.~\ref{fig:cumulative_mstar_allz_MUV}, we perform an analogous comparison for galaxies selected by dust attenuated rest-frame UV luminosity at the epoch of observation, grouping objects in bins of $M_\text{UV}$ with $\Delta M_\text{UV}=1.0$. Because $M_\text{UV}$ more directly traces recent star formation activity, this representation is useful for connecting SFH to the UV-selected galaxy populations probed by \textit{JWST}. The figure shows strong redshift evolution in the SFH at fixed $M_\text{UV}$, similar to the trends seen with $M_*$. However, $t_{50}$ and $t_{90}$ are broadly similar across $M_\text{UV}$ bins, indicating that these cumulative SF timescales are not very sensitive to the most recent episodes of star formation that dominate $M_{\rm UV}$. 

We highlight that the scatter in the cumulative stellar mass assembly histories, marked by the shaded regions, is significantly larger for galaxies observed at lower redshift. This is because lower-redshift galaxies have physically more time to form stars, allowing systems that reach similar terminal stellar masses or luminosities to do so over a wider range of growth trajectories. In contrast, galaxies forming in the ultra-high-redshift Universe exhibit a much tighter correspondence between their stellar-mass assembly histories and their terminal masses.

Considering Fig.~\ref{fig:cumulative_mstar_allz_MUV} and Fig.~\ref{fig:cumulative_mstar_allz} side by side highlights that the cumulative SF histories for $M_{\rm UV}$-controlled samples exhibit overall less scatter than those for the $M_*$-controlled samples. This is expected as a result of the known tight connection between recent SFR and $M_{\rm UV}$, since rest-frame UV luminosity traces the young stellar populations that dominate the most recent phases of growth.

These comparisons together illustrate that the bulk population of galaxies at $z \lesssim 10$ are not generally good analogues of ultra-$z$ galaxies in terms of their star formation histories and timescales. However, in rare cases, some extremely rapidly assembling systems (e.g. those well below the 16th percentile envelope in Fig.~\ref{fig:cumulative_mstar_z6}) can exhibit SFH that approach those of their ultra-$z$ counterparts.

\begin{figure}
	\centering
	\includegraphics[width=\columnwidth]{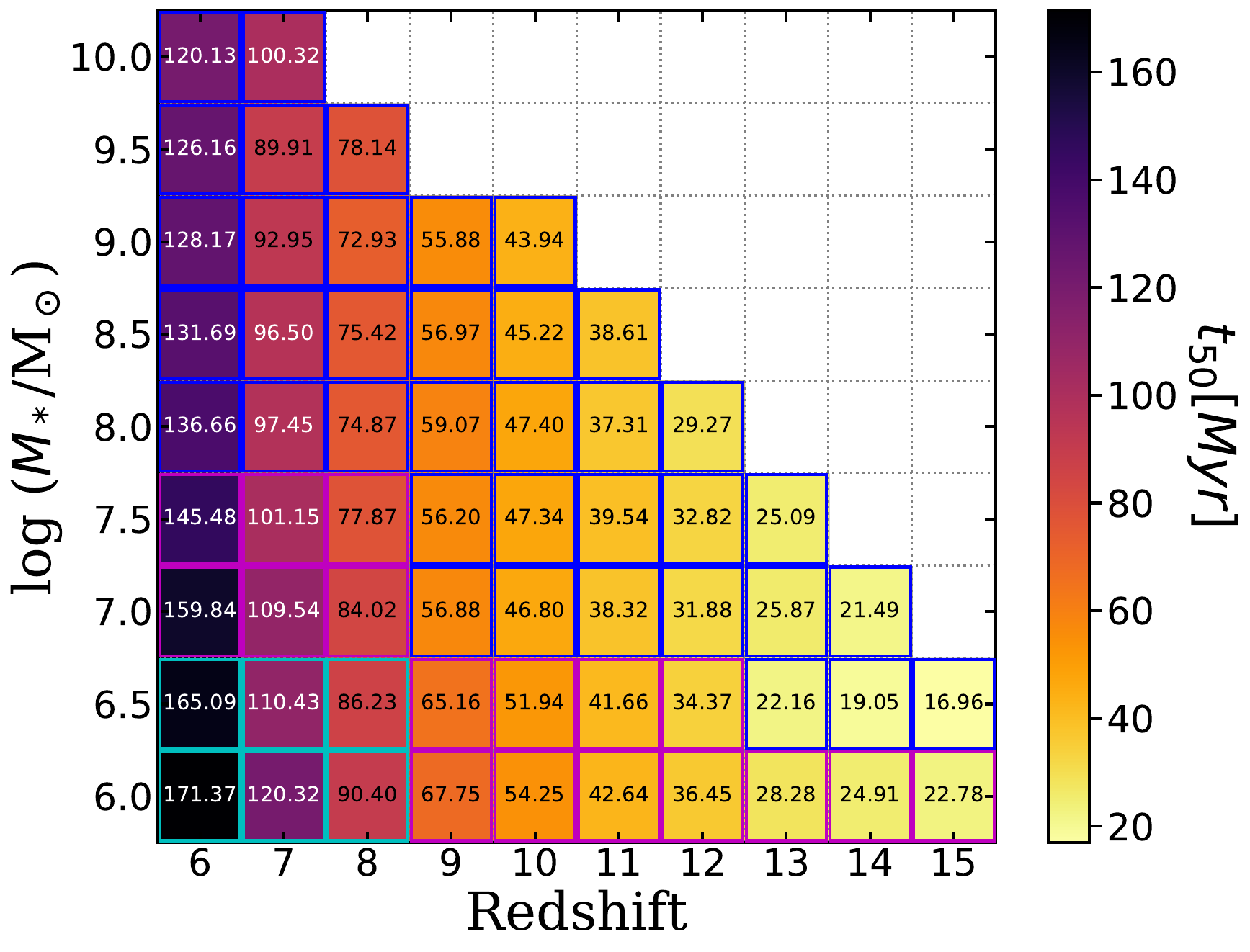}
	\includegraphics[width=\columnwidth]{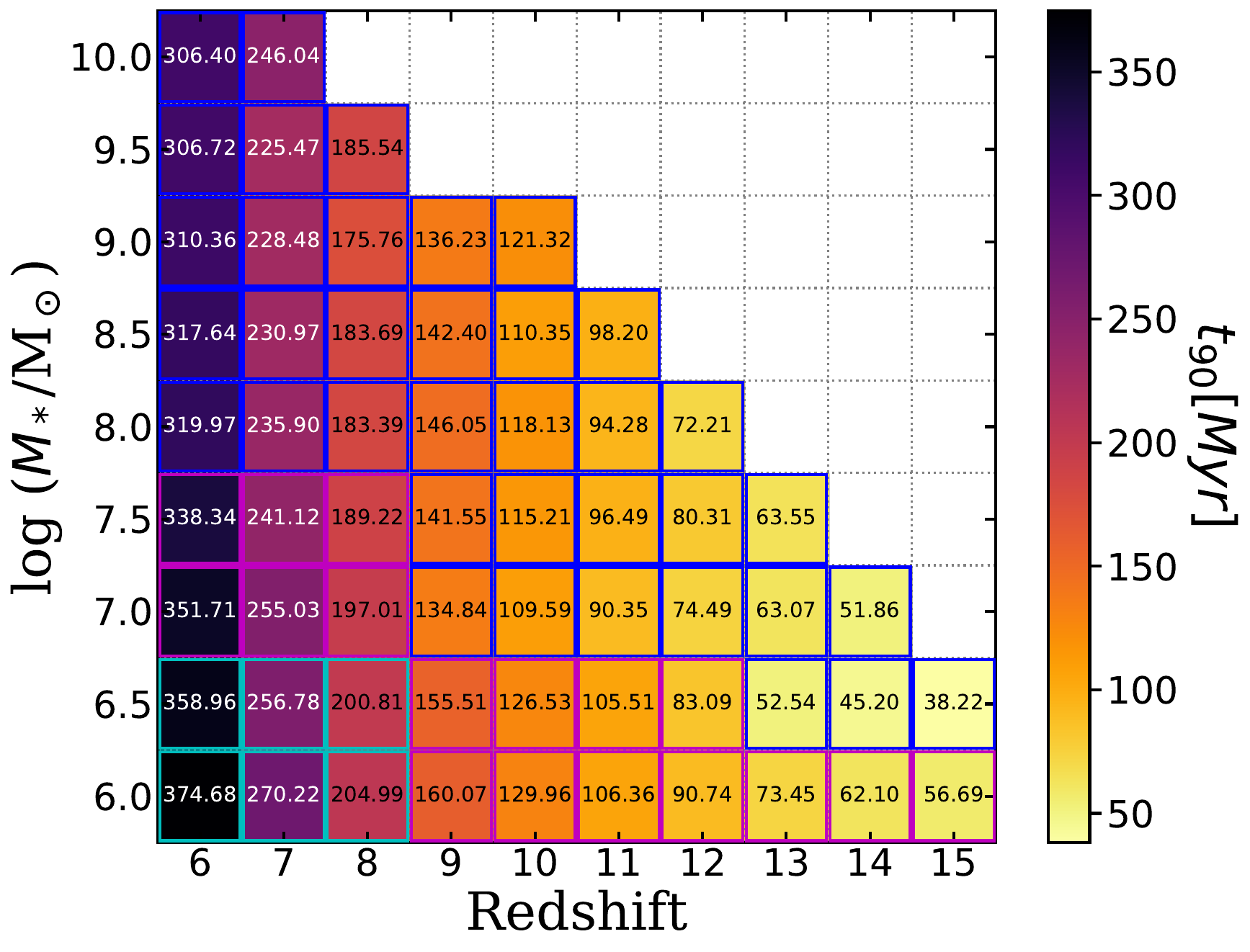}
	\caption{
		Heatmaps showing the median $t_{50}$ (\textit{top}) and $t_{90}$ (\textit{bottom}) as a function of stellar mass and redshift, where $t_X$ is the lookback time (relative to the terminal redshift) over which galaxies formed their most recent $X$ per cent of stellar mass.
        The outline colour of each grid cell indicates the \gureft\ volume from which the galaxy sample is drawn: blue, magenta, and cyan correspond to \gureft-90, \gureft-35, and \gureft-15, respectively. We find that these timescales are significantly shorter for high-redshift galaxies compared to their low-redshift counterparts. At fixed redshift, we find both star formation timescales exhibit a mild dependence on stellar mass.
	}
	\label{fig:heatmap}
\end{figure}

Following the same procedures used to construct star formation histories in Section \ref{sec:smah}, we compute $t_{50}$ and $t_{90}$ for galaxies over $6 \lesssim z \lesssim 10$ and $6 \lesssim \log(M_\ast/{\rm M_\odot}) \lesssim 10$, and summarize the results as heatmaps in Fig.~\ref{fig:heatmap}.
These heatmaps provide an overview of how the characteristic formation time-scales evolve across cosmic time and stellar mass.

We find that overall both $t_{50}$ and $t_{90}$ show a strong dependence on the output redshift, as these formation timescales are significantly shorter for high-redshift galaxies compared to their low-redshift counterparts. At fixed redshift, we find both star formation timescales exhibit a mild dependence on stellar mass, with more massive galaxies typically assembling their stellar mass over shorter intervals than lower-mass systems. These results further support that resolving the youngest stellar populations is essential for modelling photometry at $z \gtrsim 12$, since a substantial fraction of the stellar mass in these systems is formed within the most recent $\lesssim 30$ Myr. We also note that most ultra-$z$ galaxies form the bulk of their stars on time-scales significantly shorter than 100 Myr, such that reporting SFRs averaged over 100 Myr \citep[e.g.][]{Yung2024} can be of limited physical relevance in this regime (see Section~\ref{sec:SFE_burstiness} for further discussion).

We also note that, consistent with the behaviour shown in Fig.~\ref{fig:cumulative_mstar_allz}, the scatter in both $t_{50}$ and $t_{90}$ grows significantly toward lower redshift. By $z \lesssim 10$, the intrinsic spread in these timescales exceeds the mass-dependent shift in their median values, indicating that galaxy-to-galaxy diversity dominates over the systematic dependence on stellar mass.


\subsection{Relative importance of young stellar populations: interplay between SFH and bursts}
\label{sec:SFE_burstiness}

\begin{figure*}
	\includegraphics[width=1.75\columnwidth]{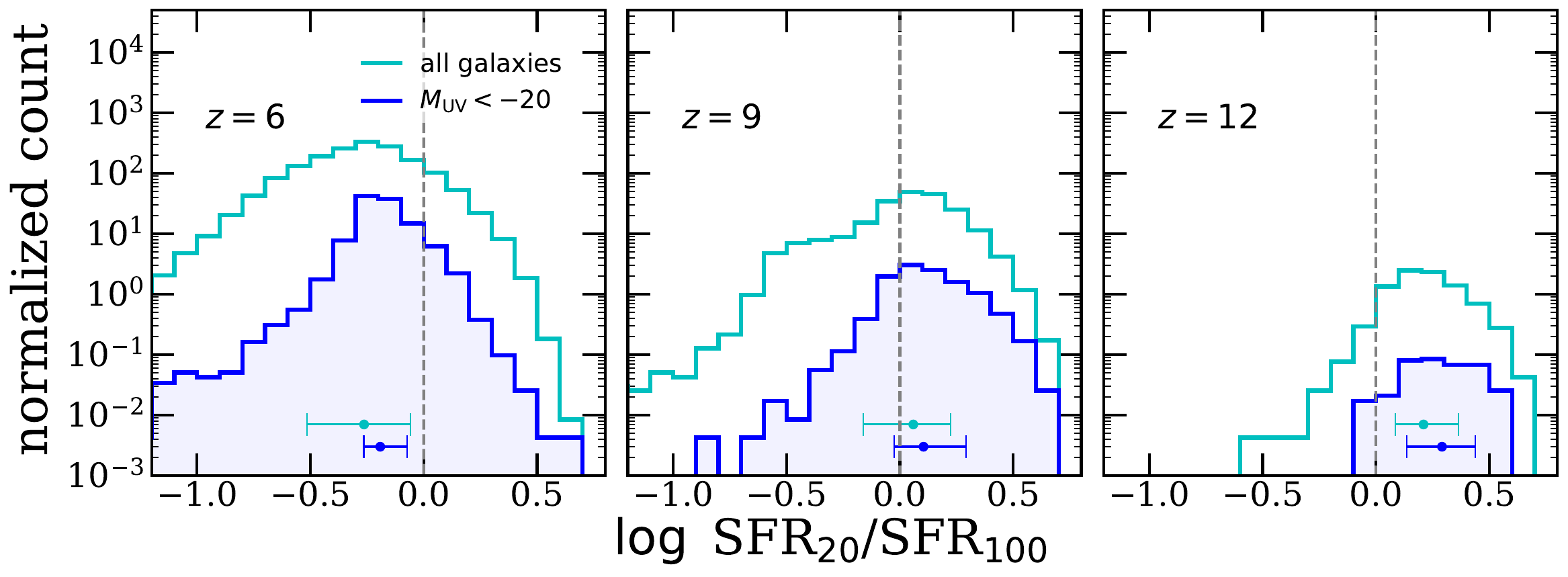}
	\caption{
		Volume-normalized distributions of the ratio of SFR$_{20}$/SFR$_{100}$ for all simulated galaxies (cyan) and for UV-luminous galaxies with $M_\text{UV} < -20$ (blue) at $z=6$ \textit{(left)}, 9 \textit{(middle)}, and 12 \textit{(right)}, simulated with VSMDPL merger trees. Data points and error bars in matching colours mark the median and the 16th to 84th percentile range of each distribution. The vertical grey dashed line marks the $\text{SFR}_{20}$=$\text{SFR}_{100}$ boundary, above which galaxies are more strongly dominated by recent star formation. 
	}
	\label{fig:SFR_ratio_hist}
\end{figure*}

\begin{figure}
	\includegraphics[width=\columnwidth]{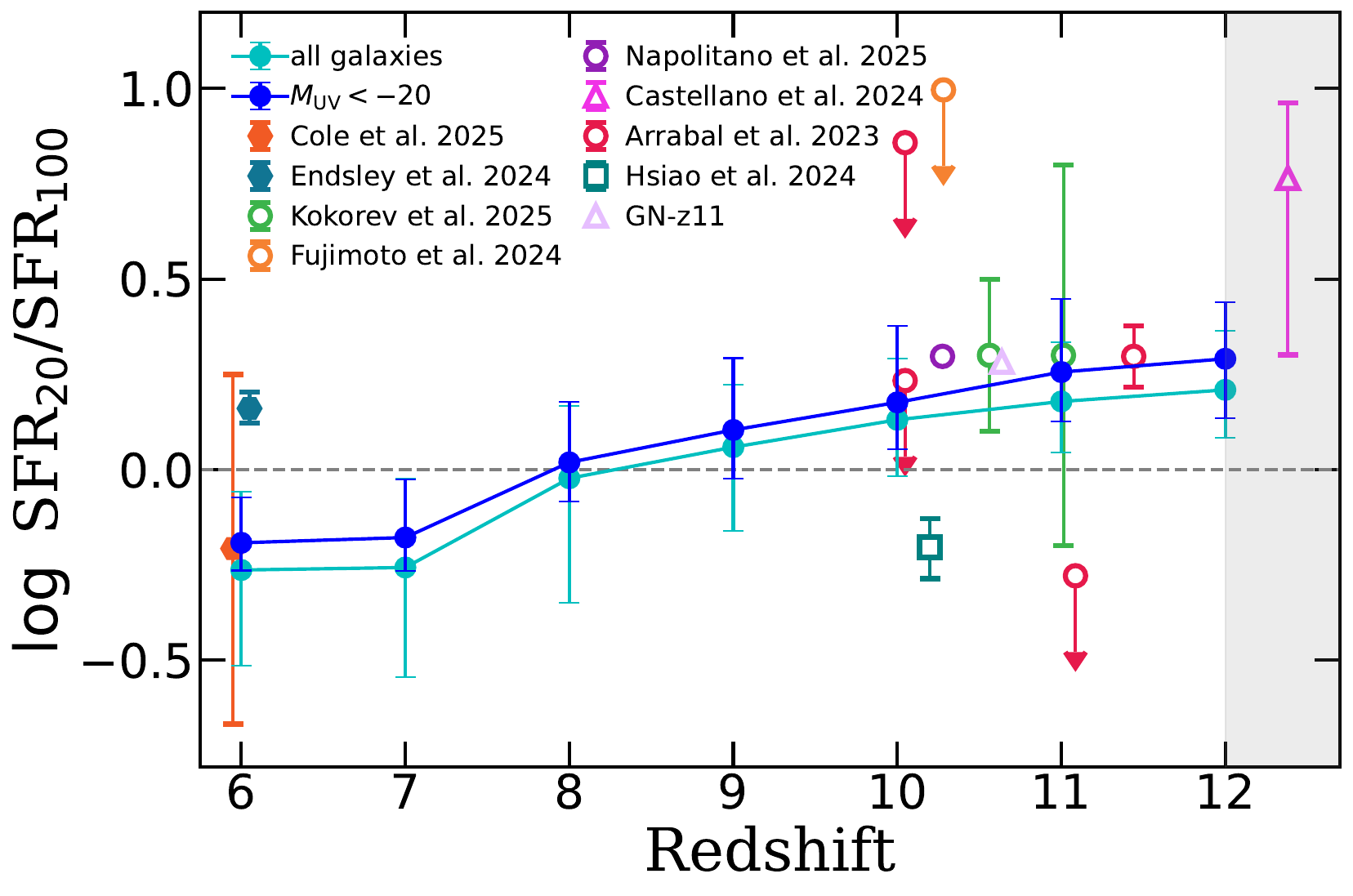}
	\caption{
		Evolution of $\text{SFR}_{20}/\text{SFR}_{100}$ over $6 \lesssim z \lesssim 12$ for all simulated galaxies (cyan) and for UV-luminous galaxies with $M_\text{UV} < -20$ (blue), simulated with VSMDPL merger trees. The data points and error bars in matching colours mark the median and the 16th to 84th percentile range of each distribution. The horizontal grey dashed line marks the $\text{SFR}_{20}$=$\text{SFR}_{100}$ boundary, above which galaxies are more strongly dominated by recent star formation. We see a clear trend of a declining median value of $\text{SFR}_{20}/\text{SFR}_{100}$ towards later cosmic times. We also show observational estimates derived from recent \textit{JWST} observations for comparison \citep{ArrabalHaro2023, ArrabalHaro2023a, Alvarez-Marquez2025, Fujimoto2024, Kokorev2025, Napolitano2025, Castellano2024, Hsiao2024,Endsley2025, Cole2025}. The observations show a similar trend, in qualitative agreement with our model predictions.
	}
	\label{fig:SFR_ratio_z}
\end{figure}

\begin{figure}
	\includegraphics[width=\columnwidth]{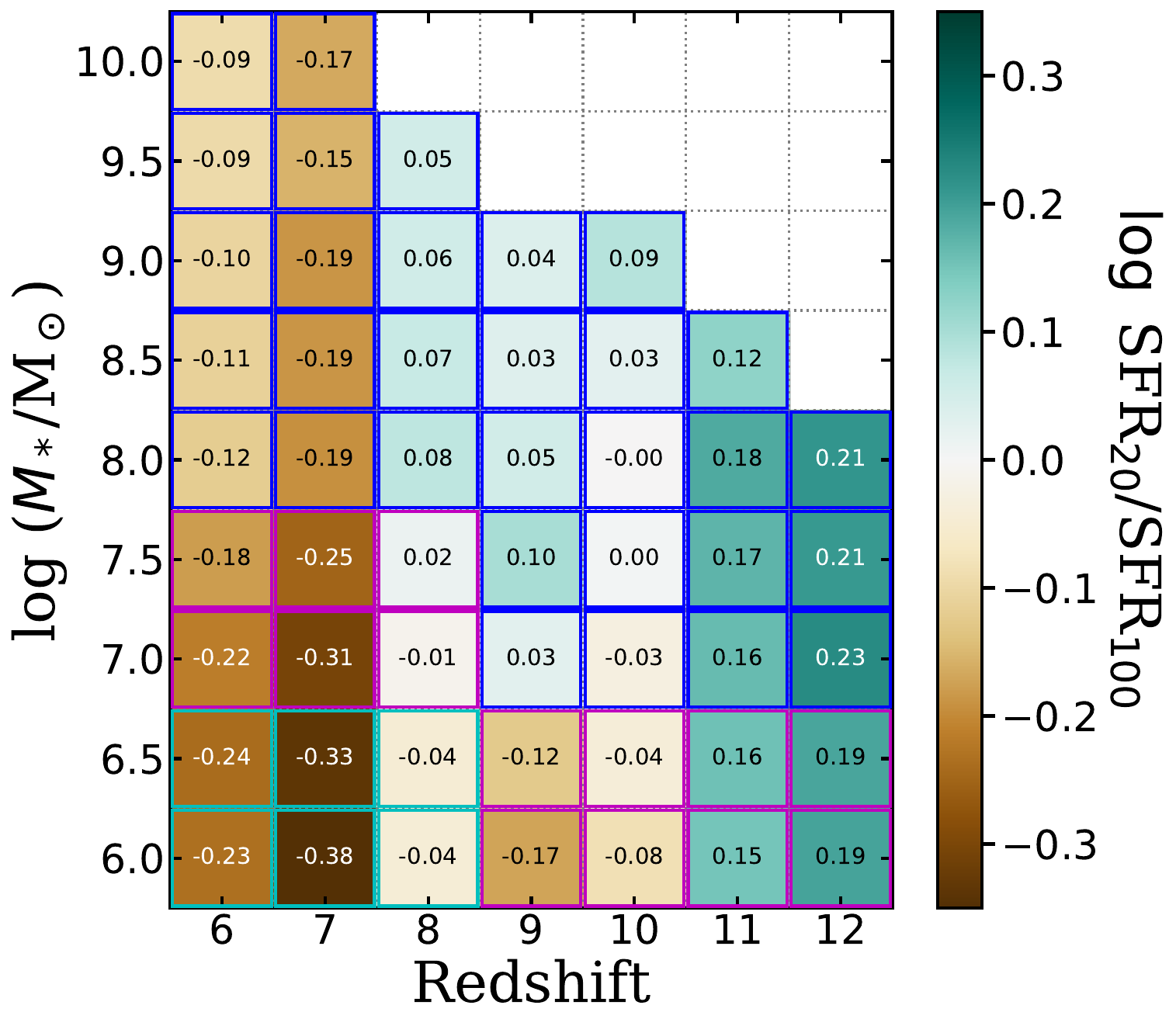}
	\caption{
		Heatmaps showing the median $\log(\text{SFR}_{20}/\text{SFR}_{100})$ as a function of stellar mass and redshift. The green end (top) of the colour bar signifies galaxy populations that are more dominated by recent star formation activity, and the brown end (bottom) of the colour bar signifies those that are more dominated by longer-term star formation activity.
        The outline colour of each grid cell indicates the \gureft\ volume from which the galaxy sample is drawn: blue, magenta, and cyan correspond to \gureft-90, \gureft-35, and \gureft-15, respectively. We find a relatively strong trend with redshift, and a weak trend with stellar mass.
	}
	\label{fig:heatmap_sfr_ratio}
\end{figure}

Ratios of SFRs averaged over shorter and longer time windows are commonly used to characterize the degree to which a galaxy is dominated by recent star formation activity, particularly during the Epoch of Reionization \citep[e.g.][and references therein]{Kokorev2025}. It is common to use observations of Balmer lines (such as H$\alpha$ or H$\beta$) to probe star formation over timescales of $\sim$ 10-20 Myr, and observations of the UV continuum to probe timescales of $\sim$ 100 Myr \citep{Erb2006, Papovich2011, Kennicutt2012}. 

In Fig.~\ref{fig:SFR_ratio_hist}, we show the volume-normalized distributions of $\text{SFR}_{20}/\text{SFR}_{100}$ for all SAM-predicted galaxies and for UV-luminous systems with $M_{\rm UV}<-20$. This magnitude threshold is chosen to approximately represent bright galaxies detectable in \textit{JWST} wide surveys at high redshift ($z\sim 6$--9) and in deep surveys at ultra-high redshift ($z\sim 12$). As already illustrated by the diversity of SFH (e.g. Figs.~\ref{fig:smah_overview} and \ref{fig:cumulative_mstar_z12}), galaxies of similar stellar mass can exhibit substantially different recent growth. This diversity naturally propagates into a broad distribution of $\text{SFR}_{20}/\text{SFR}_{100}$ values. We overplot the median and the 16th to 84th percentile range for each distribution, and mark the $\text{SFR}_{20}=\text{SFR}_{100}$ boundary for reference. Systems with $\text{SFR}_{20}/\text{SFR}_{100}> 1$ are those that are more strongly dominated by recent star formation. With the $M_{\rm UV}<-20$ selection applied, the resulting luminous subsample is preferentially weighted toward systems whose UV output is more strongly dominated by recent star formation.

In Fig.~\ref{fig:SFR_ratio_z}, we repeat this exercise across $6\leq z\leq 12$ and show the evolution of the median and the 16th to 84th percentile range of $\text{SFR}_{20}/\text{SFR}_{100}$ for both the full population and the UV-luminous subsample. The overall trend indicates that galaxies become less dominated by very recent star formation toward lower redshift. We emphasize that the distribution remains broad at all epochs, both due to the previously discussed diversity in galaxy star formation histories as well as because an individual galaxy’s SFR fluctuates over time.

Observational constraints on SFR ratios derived from a compilation of recent \textit{JWST} observations from $6 \lesssim z \lesssim 12$ are shown to be in broad agreement with our model predictions, and show hints of a trend towards higher values of $\text{SFR}_{10}/\text{SFR}_{100}$ at higher redshifts \citep{ArrabalHaro2023, ArrabalHaro2023a, Endsley2025, Cole2025, Alvarez-Marquez2025, Fujimoto2024, Kokorev2025, Napolitano2025, Castellano2024, Hsiao2024}.

Our predicted values of $\text{SFR}_{20}/\text{SFR}_{100}$ are further broken down by stellar mass and shown as a heatmap in Fig.~\ref{fig:heatmap_sfr_ratio}. 

These SFR ratios such as $\text{SFR}_{20}/\text{SFR}_{100}$ are often interpreted as indicators of the burstyness or stochasticity of star formation \citep[e.g.][]{Endsley2025}. The observed trend towards higher $\text{SFR}_{20}/\text{SFR}_{100}$ values at higher redshift could therefore be interpreted as implying either that galaxies are more bursty at earlier epochs, or that selection effects are causing us to preferentially select galaxies that are in a bursting state at higher redshift, as ''lulling'' galaxies may be too faint to detect in current observational samples. However, our results suggest a different interpretation. 

\begin{figure}
	\includegraphics[width=\columnwidth]{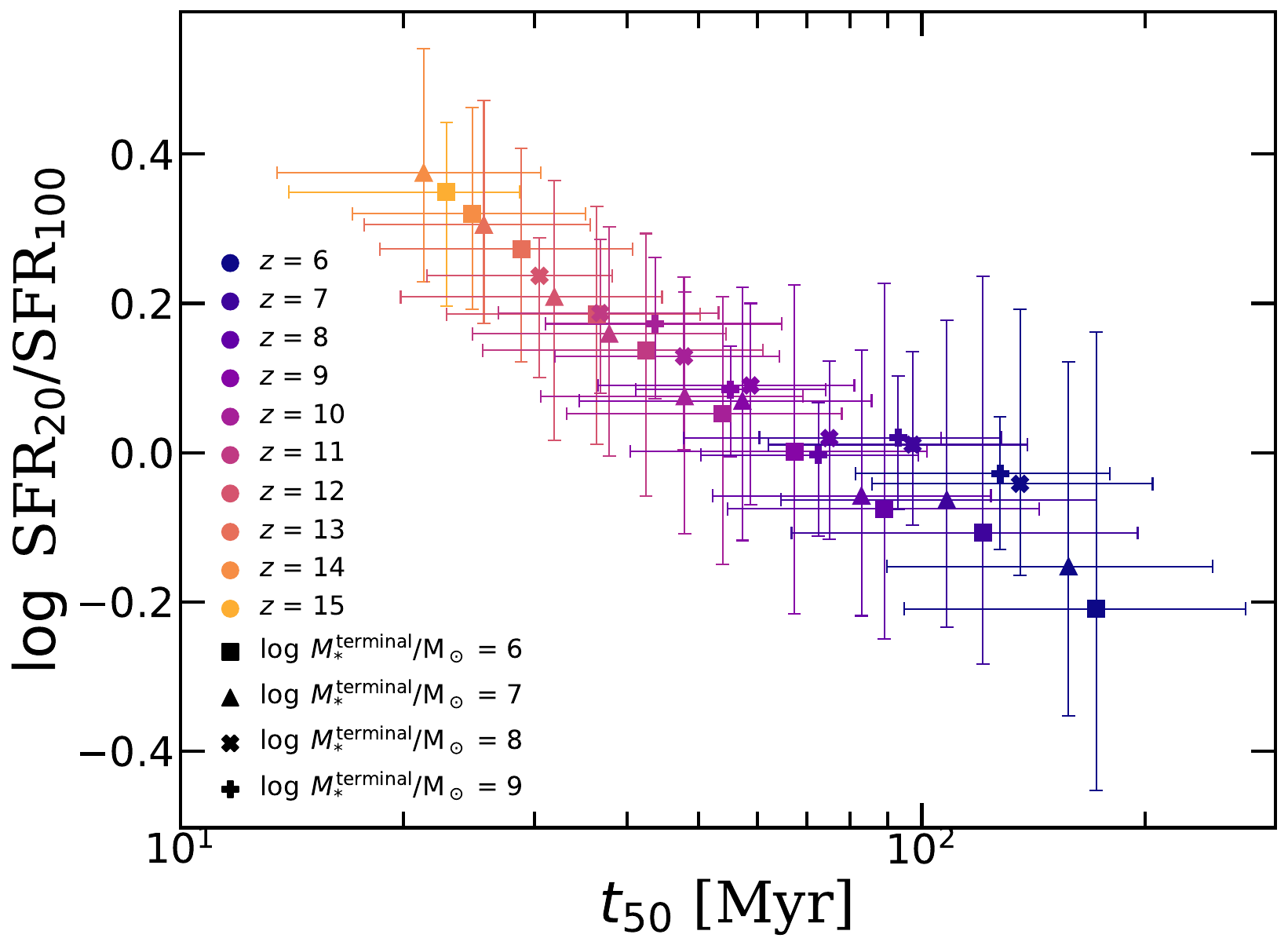}
	\caption{
		The correlation between $t_\text{50}$ and the SFR ratio SFR$_\text{20}$/SFR$_\text{100}$ computed for galaxies between $6 \lesssim z \lesssim 15$ and with $6 \lesssim \log(M_*/\text{M}_\odot) \lesssim 9$. The error bars represent the 16th and 84th percentile range in $t_\text{50}$ and SFR$_\text{20}$/SFR$_\text{100}$ among the galaxy populations. This illustrates that the decline in SFR$_\text{20}$/SFR$_\text{100}$ with increasing cosmic time may arise from the ubiquitously rising SFH coupled with lengthening characteristic star formation timescales, rather than changing star formation stochasticity (see text).
	}
	\label{fig:t50_RSFR}
\end{figure}

In Fig.~\ref{fig:t50_RSFR}, we show the relation between $t_{50}$ and the SFR ratio $\text{SFR}_{20}/\text{SFR}_{100}$ for model galaxies over $6 \lesssim z \lesssim 15$ and $6 \lesssim \log(M_*/\text{M}_\odot) \lesssim 9$. The error bars indicate the 16th to 84th percentile ranges in both $t_{50}$ and $\text{SFR}_{20}/\text{SFR}_{100}$. These quantities are computed from the median and the 16th to 84th percentile summary stellar-mass growth histories introduced in Section~\ref{sec:smah} and further described in Appendix~\ref{app:SMAH_fitting_function}. They therefore characterize the overall growth of the galaxy population, while short-timescale fluctuations in individual galaxies are smoothed by construction. The tight correlation between these quantities indicates that the predicted redshift evolution for our model galaxies shown in Fig.~\ref{fig:SFR_ratio_z} is largely driven by the systematic shortening of characteristic galaxy growth time-scales toward earlier cosmic times, \emph{not by star formation stochasticity}.

Thus, while $\text{SFR}_{20}/\text{SFR}_{100}$ may provide a useful burstiness diagnostic for galaxies whose SFHs extend over many hundreds of Myr or longer, as is typical for local and low-redshift systems, its interpretation becomes less straightforward at high and ultra-high redshift. In this regime, the available cosmic time for star formation is itself limited, and the ratio is strongly influenced by the compressed global growth history rather than only by recent stochastic bursts. In this sense, many ultra-$z$ galaxies may appear `bursty' by low-redshift standards simply because their entire SFHs unfold over timescales comparable to what would be classified as a single burst in nearby galaxies. Some of these systems may therefore be observed during the rapid assembly phase of their first major episode of star formation, rather than during a burst superimposed on a long underlying SFH.

\section{Discussion} 
\label{sec:discussion}

\subsection{The impact of ubiquitous young-star-dominated star formation histories}

The main result of this work is that the stellar populations of $z\gtrsim 6$, and even more so $z\gtrsim 10$, galaxies are heavily dominated by very young stars. The most recent 50\% of the stellar mass in $z\sim 10$ galaxies was formed over the past $\sim 55$--60 Myr, and the most recent 90\% was formed over $\sim 110$--120 Myr (see Fig.~\ref{fig:heatmap}). This result has several important implications for modelling and interpreting observations of ultra-high-z galaxies.

\subsubsection{Implications for forward modelling}
As illustrated in Figs.~\ref{fig:MUV_age} and \ref{fig:SED_decomposition}, the FUV luminosity of young stellar populations can evolve extremely rapidly over the first $\sim$20 to 30 Myr and can contribute significant fraction (as high as $\gtrsim90$ percent), of an ultra-high-redshift galaxy's UV continuum emission. Properly accounting for the contribution from these young stellar populations is therefore crucial to computing galaxy rest-frame UV and optical luminosities, as well as their observed-frame near-infrared magnitudes. This is illustrated in Fig.~\ref{fig:UVLF_8panels_compY24}, which shows that adopting a finer age binning in the SFH-to-SED mapping yields UV magnitudes that are up to $\sim 2$ mag more luminous than those obtained with the approach adopted in \citet{Yung2024}, where the central age within a 10 Myr wide bin was used.

In large volume cosmological hydrodynamical simulations, each star particle typically represents an unresolved stellar population rather than an individual star, so finite mass resolution inevitably coarsens the sampling of recent star formation. When star particles are too massive, the youngest stellar populations, especially those with ages of only a few Myr, may be poorly represented or even absent in individual galaxies or snapshots, despite ongoing star formation, simply because the formation of one additional star particle corresponds to a comparatively large amount of stellar mass. This matters because the emergent UV and nebular output is strongly weighted toward very young massive stars: H$\alpha$ traces stars younger than roughly 5–10 Myr, and even the far-UV can be dominated by substantially younger populations, as illustrated in this work. Several simulation pipelines therefore introduce explicit sub-resolution corrections for this problem. For example, the EAGLE+\textsc{skirt} post-processing framework re-samples star-forming gas and the youngest stellar particles into sub-grid star-forming regions, with newly formed components younger than 10 Myr treated separately to recover the luminosity contribution of embedded young sources \citep{Camps2016}. More generally, work on stochastic IMF sampling and low-mass galaxy simulations shows that limited sampling of the high-mass stellar population can change feedback, star-formation histories, and H$\alpha$/FUV-like observables, while studies of high-redshift UV luminosity functions have noted that the faint-end turnover can reflect the inability of a given mass resolution to capture recent star formation histories adequately \citep{Applebaum2020, Applebaum2021}. Thus, unless some re-sampling or sub-grid correction is applied, low-resolution simulations can systematically underestimate UV luminosities—particularly for low-mass, bursty, or high-redshift galaxies.

The hard ionizing radiation from these young stellar populations is also expected to illuminate the ISM and significantly increase the contribution from nebular continuum and strong emission lines \citep[e.g.][]{Izotov1999, Wilkins2013, Hirschmann2017, Hirschmann2019}, which may have a strong impact on observed-frame photometry and colours, especially in medium and narrow bands \citep[e.g.][]{Wilkins2022, Hirschmann2023, Scharre2024}. We plan to explore this systematically in a future work, where we implement a nebular emission model similar to that presented by \citet{Hirschmann2017, Hirschmann2019, Hirschmann2023} within the Santa Cruz SAM (Yung et al. in preparation). 

\subsubsection{Implications for interpreting observations and SED fitting}
Similarly, the generically strongly rising SFH and short expected timescales for galaxy assembly in the ultra-high-z universe have implications for interpreting JWST observations, in particular for estimating physical properties via SED fitting. Although rest-UV selected samples are inevitably biased towards galaxies with vigorous recent star formation, our results suggest that this does not \emph{necessarily} imply that samples are biased towards galaxies experiencing a very short-lived burst of star formation. Moreover, our results can be used to inform SFH priors used in SED fitting. Assumptions of constant or declining SFH are likely to yield unphysical and biased results. Similarly, the use of coarse age bins (which then implicitly imposes a constant SFH over the duration of the bin) can significantly affect the physical parameters derived from SED fitting. If the age bins are too broad, the luminosity from very young stars may be mapped onto older stellar populations that are intrinsically fainter at UV wavelengths. As a result, the inferred stellar mass needed to reproduce the observed flux may be biased high, and the corresponding recent SFR may likewise be overestimated \citep{Sun2023, Tacchella2023, Haskell2024}. The functional form describing the cumulative SFH presented in Eqn.~\ref{eqn:sfh} can be used as a template prior for SED fitting.

\subsection{Downsizing at Cosmic Dawn}
\label{sec:SFE_evolution}

We have shown that massive (and, to a lesser extent, UV-luminous) galaxies identified at $6 \lesssim z \lesssim 12$ form \emph{earlier and more rapidly} than their lower mass counterparts (Fig.~\ref{fig:cumulative_mstar_z6}, \ref{fig:cumulative_mstar_allz}, \ref{fig:heatmap}). This behaviour, also qualitatively seen in the lower redshift Universe, is sometimes referred to as ``downsizing''. However, the physical origin of this behaviour is likely somewhat different at these very early epochs. Downsizing at $z\lesssim 2$ is likely due in part to the earlier quenching of more massive galaxies, presumably by black hole feedback. Galaxies at cosmic dawn remain rapidly star forming, but their star formation timescales are modulated by the formation histories of the underlying dark matter halos as well as by the star formation efficiencies within those halos. More massive dark matter halos actually form \emph{later} than less massive counterparts identified at a given cosmic epoch, in terms of the time when they had assembled a given fraction of their final mass (e.g. $t_{50}$). However, if we consider a fixed lookback time, more massive halos also tend to have progenitors that are more massive \emph{in an absolute sense} than those of less massive halos. 

In our models, the onset of star formation is largely determined by the time when a halo becomes massive enough to cool via atomic processes ($T_{\rm vir}> 10^4$K), since we do not include molecular cooling or metal cooling below $10^4$ K. Massive halos are more likely to have a progenitor that crosses the atomic cooling limit. Moreover, as discussed briefly in Section~\ref{sec:results:SFE}, the galaxy-scale star formation efficiency in our models has a strong dependence on the halo circular velocity, through the parameterization of the mass loading of stellar driven galactic winds. This represents the greater difficulty that supernovae may have in ejecting gas from deeper potential wells. In addition, the super-linear slope of the Kennicutt–Schmidt SF relation adopted in the models presented in this work \citep{Schmidt1959, Schmidt1963, Kennicutt1989, Kennicutt1998} leads to higher SFE at fixed halo virial velocity at higher redshift, where the ISM is denser on average \citep{Somerville2015,Yung2019a,Somerville2025}. As shown in Fig.~\ref{fig:SFE_evolution}, these combined factors lead to SFE that increase systematically as halos grow over cosmic time.

Even so, the models presented here do not reproduce the observed UVLF at $z\gtrsim 12$ (Fig.~\ref{fig:UVLF_8panels_ext}), suggesting that additional or modified physical processes may be needed to explain the $z\gtrsim 12$ observations. \citet{Yung2025} similarly found that an evolving SFE is required in order to empirically reproduce the observed UV LFs. Many possible solutions to this ``early bright galaxy excess'' have been suggested in the literature (see the extensive discussion and references in \citet{Somerville2025} and \citet{Somerville2026}). In \citet{Somerville2025}, we used a similar framework to the models presented here, but incorporated a ``density modulated star formation efficiency'' model motivated by detailed simulations on giant molecular cloud scales. This model predicts more rapid and pronounced evolution in the SFE due to the higher ISM gas densities in the early Universe, and is able to match the observed $z\gtrsim 12$ UVLFs. Were we to repeat the measurements performed here with the DMSFE model, we expect that some of the quantitative results of this analysis would change, in particular, star formation timescales would be slightly shifted towards earlier times. However, we would not expect any of the qualitative conclusions presented here to change in the context of the DMSFE framework. 

At fixed halo mass and redshift, we find significant scatter in the predicted formation times. This likely reflects differences in halo assembly history and large-scale environment, with galaxies in overdense regions expected to collapse earlier and reach high gas surface densities sooner. Such systems may therefore enter efficient star-forming phases earlier, potentially contributing to the dispersion in SFE at fixed mass. This also raises the possibility that observed \textit{JWST} samples are biased toward overdense regions, where early-forming, UV-bright galaxies are preferentially found \citep{Jespersen2022, Jespersen2025, Weaver2023, Weibel2024}.

\subsection{Caveats and limitations of our study}

\subsubsection{Short timescale star formation stochasticity}
\label{sed:caveats_bursts}

\citet{Iyer2020} presented an extensive discussion of the broad range of physical processes that can strongly modulate star formation on short timescales, thereby producing episodic or bursty star formation histories. The Santa Cruz SAM includes several mechanisms expected to contribute to such variability, most notably galaxy--galaxy mergers and merger-induced starbursts, whose effective timescales should become significantly shorter in the ultra-high-redshift Universe than at low redshift. More generally, the Santa Cruz SAM also captures processes that regulate star formation over longer intervals of tens to hundreds of Myr, including galactic-scale stellar feedback and galaxy and halo-scale baryon-cycling processes. However, a variety of processes acting on giant molecular cloud (GMC) scales are expected to lead to star formation stochasticity on $\lesssim 10$ Myr timescales \citep[e.g.][]{Leitherer1999, Tan2000, Tasker2011, Faucher-Giguere2018, Benincasa2019}. In its current configuration, the Santa Cruz SAM treats star formation as a galaxy-averaged process: the Kennicutt--Schmidt-like star formation prescription effectively absorbs the averaged impact of these sub-grid processes, but does not explicitly resolve the stochastic formation, disruption, and feedback-regulated cycling of individual dense clouds. 

While we showed that an observed increasing SFR$_{20}$/SFR$_{100}$ ratio toward higher redshift does not necessarily imply enhanced burstiness, and can instead arise from systematically shorter star formation timescales, we do not rule out the possibility that genuinely bursty star formation affects the SFHs of ultra-high-$z$ galaxies and their observational signatures. Indeed, both numerical simulations and empirical modelling suggest that burstiness may be enhanced in lower-mass haloes, which increasingly dominate the galaxy population at earlier times \citep{Gelli2024, Munoz2026}. We therefore regard short timescale SF stochasticity as an important remaining uncertainty in the characterization of galaxy SFH, but one that should be constrained jointly by luminosity functions, emission-line to UV diagnostics, and clustering statistics rather than inferred from $\mathrm{SFR}_{20}/\mathrm{SFR}_{100}$ alone.

\subsubsection{Limited dynamic range and incompleteness in progenitor populations}
\label{sed:caveats_dynamicrange}

We acknowledge that the finite dynamic range of individual \gureft\ volumes can affect the earliest stages of the predicted SFHs, particularly at cosmic time $\lesssim 400$ Myr, as shown in the bottom row of Fig.~\ref{fig:SFH_mstar}. This limitation is most apparent for merger trees drawn from \gureft-90, the largest and lowest-resolution box, where progenitor haloes can fall below the mass-resolution limit and cause the trees to be prematurely truncated.

However, the normalized assembly histories in Figs.~\ref{fig:cumulative_mstar_z6} and \ref{fig:cumulative_mstar_z12} show that most galaxies do not form a substantial fraction of their stellar mass in this earliest regime. In many cases, even well-resolved merger trees exhibit median growth histories that predict zero star formation at cosmic time $\lesssim 400$ Myr. Thus, while progenitor incompleteness can affect the inferred earliest star-forming episodes, it is unlikely to dominate the characteristic timescales $t_{50}$ and $t_{90}$ reported in this work. Future work will address this limitation by extending merger trees using the generative machine learning based approach \textsc{florah} presented in \citet{Nguyen2024, Nguyen2025}.

A related subtlety is how to treat zero-SFR entries at the earliest times when constructing population-level median SFHs. These zeros can arise from two physically distinct situations: genuinely late-forming galaxies whose star formation has not yet begun, and early-forming galaxies whose progenitor branches have been prematurely truncated by the mass-resolution limit. In practice, these cases are difficult to distinguish robustly from the merger trees alone, since there is no unambiguous criterion for determining whether a tree has physically reached its earliest progenitor or has instead fallen below the resolution limit. Masking all zero-SFR entries would therefore preferentially remove genuinely late-forming systems and bias the median SFH high at early times. We therefore retain these zeros when computing median SFHs, while cautioning that the resulting early-time medians should be interpreted as conservative estimates in regimes where progenitor incompleteness may be important.


\section{Summary and Conclusions}
\label{sec:summary}

In this work, we carry out an in-depth investigation of the star formation histories of galaxies from Cosmic Dawn through the Epoch of Reionization, leveraging the physically motivated Santa Cruz SAM and the dark matter halo merger trees extracted from the \gureft\ suite of cosmological simulations. We also present updated UV LF predictions after adopting finer age bins of $\Delta \log(\text{age}) = 0.1$ for the SFH-to-SED mapping, which are substantially narrower than the fixed 10 Myr bins adopted in previous work \citep{Yung2024}.  We reach the following main conclusions:

\begin{itemize}

    \item  We introduce an updated SFH to SED mapping that re-grids the SAM-predicted SFHs onto the finer age bins adopted by the \textsc{bpass} stellar population synthesis models, and construct composite stellar continuum SEDs that properly account for the contributions from very young stellar populations with ages $<10$ Myr.

    \item We show that this improvement can yield a $\sim$2 mag increase in far-UV magnitude relative to previous results that used much coarser 10 Myr age bins, producing UV luminosity functions that are in substantially better agreement with \textit{JWST} observations up to $z\sim12$. This significantly reduces the discrepancy between observations and simulations without invoking new prescriptions for star formation and stellar feedback.
	
    \item We present updated photometry and UV luminosity functions that incorporate rarer, more massive galaxies by supplementing \gureft\ with merger trees of massive halos drawn from the VSMDPL cosmological simulation. This provides more reliable predictions for the bright end of the UVLF, reaching number densities as low as $\lesssim 10^{-6}\,{\rm mag^{-1}\,Mpc^{-3}}$, compared to runs based on \gureft\ merger trees alone, which is necessarily for reproducing some of the rarest, most luminous sources detected by \textit{JWST}.

    \item Galaxies at $z \gtrsim 6$ have \emph{rapidly rising} star formation histories on average. We present a functional form that can be used as a template for SFH priors in SED fitting.

    \item Progenitor-descendant mappings are intrinsically \textit{broad}, that is, galaxies with similar terminal stellar masses can have diverse assembly histories, and conversely ultra-high-$z$ galaxies of similar mass can yield descendants spanning a wide range of stellar masses at later times.

    \item Ultra-high-$z$ galaxies assemble their stellar masses on strongly compressed time-scales compared to their lower-redshift counterparts. For galaxies observed at $z\gtrsim12$, the typical time required to form the most recent 50\% (90\%) of their stellar mass is $t_{50} \lesssim 30$ Myr ($t_{90} \lesssim 70$ Myr), a factor of $\sim$3--4 shorter than for comparable galaxies near the end of the EoR ($z\sim6$).

    \item We show that typical ultra-high-$z$ galaxies have characteristic star-formation timescales that are significantly shorter than 100 Myr, suggesting that their UV luminosities are not in fact good indicators of star formation on this timescale.
    
    \item We find that the decrease of the $\text{SFR}_{20}$/$\text{SFR}_{100}$ ratio from $z\sim 12$ to $z\sim 6$, as suggested by recent \textit{JWST} observations, may be largely driven by ubiquitously rising SFH coupled with lengthening characteristic star formation timescales towards later epochs, and is not necessarily indicative of decreasing burstiness.

\end{itemize}

\section*{Acknowledgements}

The analysis in this work was carried out with \textsc{astropy} \citep{Robitaille2013, Price-Whelan2018}, \textsc{pandas} \citep{Reback2022}, \textsc{numpy} \citep{vanderWalt2011}, and \textsc{scipy} \citep{Virtanen2020}. \texttt{pathfinder} \citep{Iyer2024} was used for literature searches. 
The GUREFT simulation suite and Santa Cruz semi-analytic galaxy formation model was run on the Flatiron Institute computing cluster rusty , managed by the Scientific Computing Core (SCC).
AY is supported by a Giacconi Fellowship from the Space Telescope Science Institute, which is operated by the Association of Universities for Research in Astronomy, Incorporated, under NASA contract HST NAS5-26555 and JWST NAS5-03127. The Flatiron Institute is supported by the Simons Foundation. 
This work was performed in part at the Aspen Center for Physics, which is supported by National Science Foundation grant PHY-2210452.
We are grateful to Raffaella Schneider, Brant Robertson, Roberto Maiolino, and Volker Bromm for organizing the Kavli Institute for Theoretical Physics (KITP) program ``Cosmic Origins: The First Billion Years'', and the KITP for hosting this program. This research was supported in part by grant NSF PHY-2309135 to the Kavli Institute for Theoretical Physics (KITP).
This work is based on observations made with the NASA/ESA/CSA James Webb Space Telescope, obtained at the Space Telescope Science Institute, which is operated by the Association of Universities for Research in Astronomy, Incorporated, under NASA contract NAS5-03127. The Flatiron Institute is supported by the Simons Foundation.

\section*{Data Availability}

The data used in this work will be made available upon request.



\bibliographystyle{mnras}
\bibliography{ultra-z-sfhist} 



\appendix

\section{Combining distribution functions across multiple simulated volumes}
\label{app:LF_combine_scheme}

\begin{figure}
    \includegraphics[width=\columnwidth]{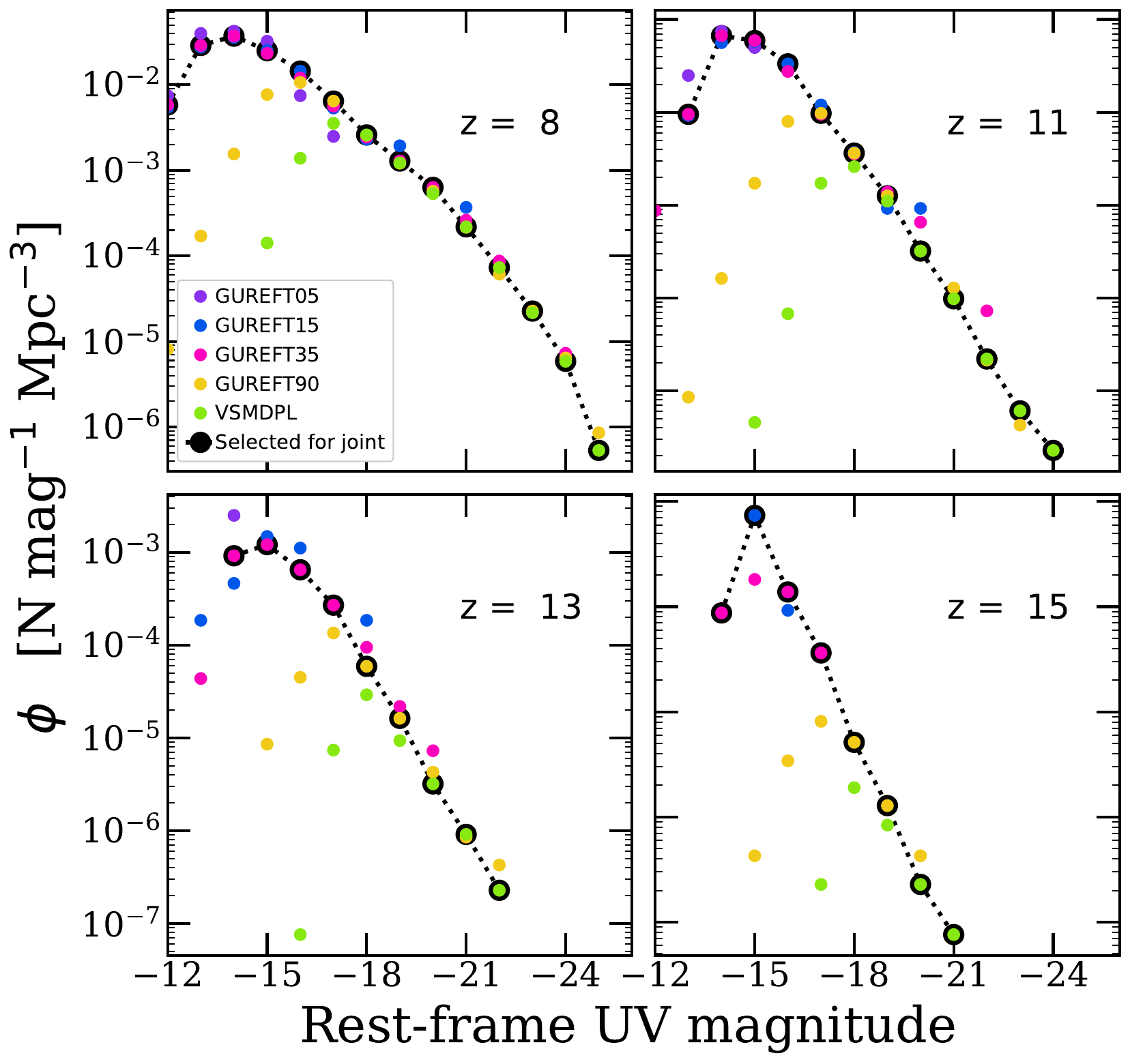}
    \caption{
        Demonstration of the distribution-function combination scheme using UVLFs from individual \gureft\ volumes and VSMDPL. High-contrast colours show UVLFs measured in each simulated volume. The data points selected for the combined UV LFs are highlighted with thick black marker borders and connected with black dotted lines. Bins deemed unreliable due to small-number statistics (too few available galaxies limited by simulated volume) or incompleteness (set by mass resolution) are excluded from the combined distribution function.
    }
    \label{fig:UVLF_joint_demo}
\end{figure}

In \citetalias{Yung2024}, distribution functions, including UV luminosity functions (UVLFs) and stellar mass functions (SMFs), from the four \gureft\ volumes were inspected and combined manually. In this work and the companion work of \citet{Somerville2025}, the number of model variants increases substantially, including the new density modulated star formation efficiency (DMSFE) model with free parameters, as well as models configured to account for dust attenuation and addition burstiness, and observed-frame luminosity functions predictions across many \textit{JWST} bands), motivating an automated and reproducible scheme for combining distribution functions across simulation volumes.

An intuitive approach would be to stitch samples across volumes by selecting galaxies based on a halo property such as $M_\text{h}$ (or $V_\text{max}$), then measuring the distribution function in the combined sample. However, because galaxy-halo relations (e.g. stellar-to-halo mass ratios, $M_\text{UV}$-$M_\text{h}$ scaling relations) exhibit significant scatter and can shift across model variants, such selection-based stitching can introduce artificial turnovers and discontinuities in the resulting distribution functions. For this reason, we combine the outputs at the \textit{distribution function level} rather than the \textit{galaxy catalogue} level.

For a given simulated volume, an individual distribution function suffers from two generic limitations:
\begin{enumerate}
    \item \textit{Incompleteness at the faint/low-mass end:} the faint (or low-mass) bins become unreliable when the contributing galaxies reside in halos near the mass-resolution limit of the simulation volume.
    \item \textit{Shot noise at the bright/high-mass end:} toward the bright (or massive) end, the number of objects per bin drops rapidly, increasing Poisson uncertainties. In this regime, a single object can spuriously inflate the inferred number density, while in other realizations the same bin may be empty.
\end{enumerate}

To mitigate these issues, we impose a minimum-occupancy threshold for each bin to be deemed reliable. Specifically, we require a minimum number of objects per bin of: 50 for $z < 11$, 35 for $11 \leq z < 13$, 25 for $13 \leq z < 14$, 4 for $14 \leq z < 15$, and 1 for $z \geq 15$. In regimes where bins are not affected by incompleteness or shot noise (i.e., where the distributions are effectively converged), the predicted distribution functions agree across volumes and the choice of volume is unimportant. In the overlap regime where two volumes both contribute but yield different number densities, we adopt the \emph{larger} of the two values. This choice preferentially rejects faint-end points from larger boxes that are susceptible to incompleteness, while retaining the better-resolved measurements from smaller boxes.

Fig.~\ref{fig:UVLF_joint_demo} illustrates this procedure by showing UVLFs from individual \gureft\ volumes and VSMDPL at $z=8, 11, 13,$ and 15, highlighting the bins selected for the combined UV LFs using the criteria above. An advantage of this approach is that it remains robust to shifts in galaxy-halo relations across model configurations (including dust and starburst implementations), since the selection is performed directly on the measured distribution functions rather than on halo-based sample cuts.


\section{Extended stellar mass functions with VSMDPL}
\label{app:extended_SMFs}

\begin{figure}
    \includegraphics[width=\columnwidth]{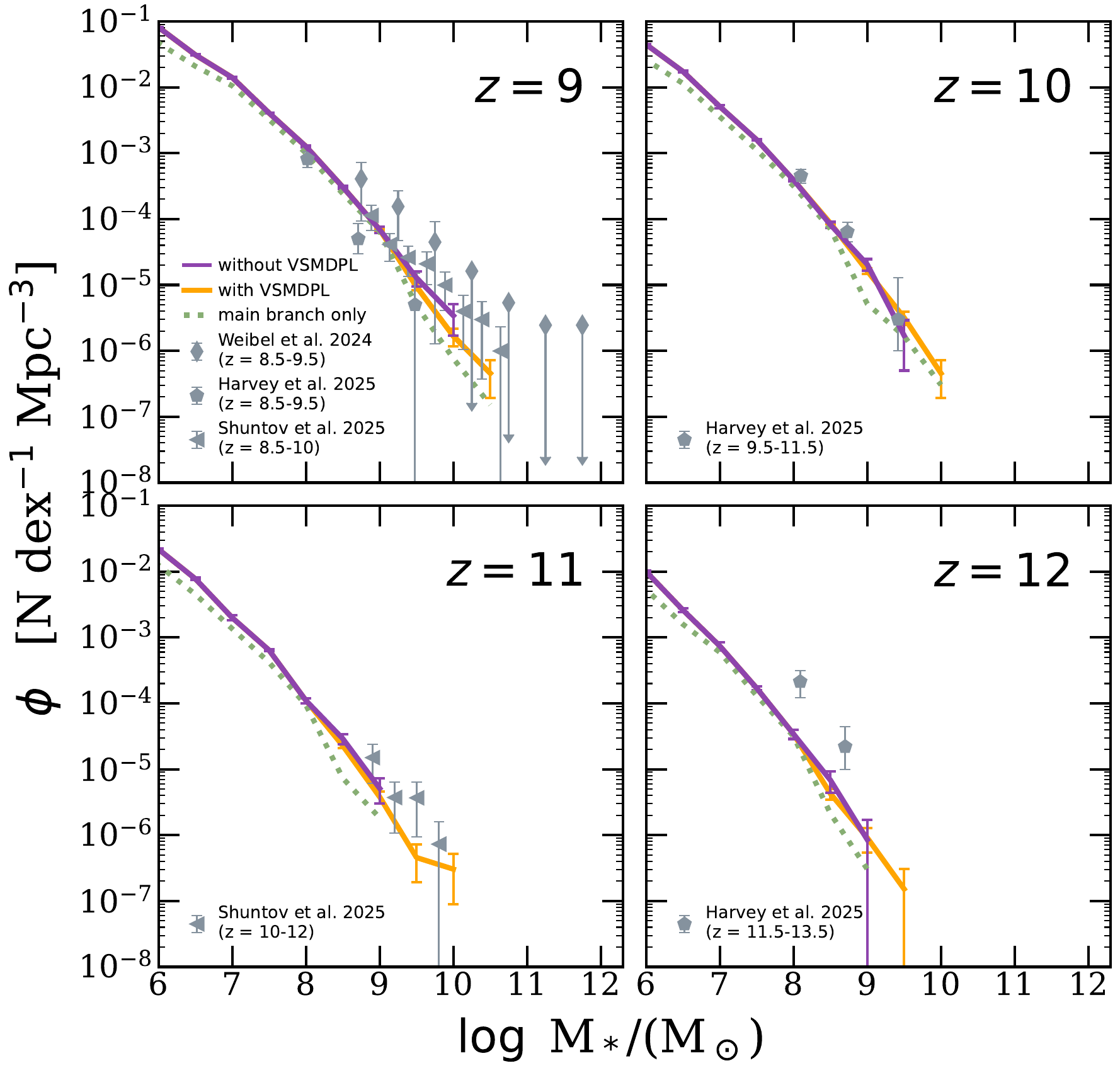}
    \caption{
        Stellar mass functions (SMFs) at $z = 9$ to 12. Symbols shown are observational constraints from \citet{Weibel2024, Harvey2025, Shuntov2025}. The purple lines show results run on \gureft\ merger trees (without VSMDPL, same as \citetalias{Yung2024}) and the orange lines show the extended SMFs that include results from VSMDPL merger trees. The error bars represent the Poisson noise associated with these simulated SMFs, estimated from the number of galaxies available in each stellar mass bin. We highlight that the most massive bins, which have significantly lower object counts compared to other bins, would likely result in an over estimated number density (see text for further discussions). In addition, we show SMFs for galaxies run only on the main branch of the merger trees from both \gureft\ and VSMDPL (green dotted).
    }
    \label{fig:SMF_ext}
\end{figure}

Fig.~\ref{fig:SMF_ext} shows high- to ultra-$z$ SMFs with extension to galaxies in more massive halos drawn from VSMPDL are provided. Tabulated SMFs are provided on our data release website.

To illustrate the benefit of including additional high-mass haloes from the larger VSMDPL volume, we show Poisson counting uncertainties on the simulated SMFs. These are estimated as $\sigma = \sqrt{N}/(V\Delta M_*) = \sqrt{\phi/(V\Delta M_*)}$, where $N$ is the number of galaxies included in mass bin with width $\Delta M_*$ sourced from a simulation with volume $V$, and the expression is modified to directly compute uncertainty for the SMF $\phi$. With the inclusion of VSMDPL, the Poisson uncertainties are noticeably reduced in the high-mass end, and the SMFs extend to stellar masses not sampled by \gureft-90 alone while retaining meaningful statistical constraints. At $z \gtrsim12$, however, the most massive bins contain very few objects, and thus the corresponding uncertainties become substantially larger.

\section{Fitting function for fraction of stars formed}
\label{app:SMAH_fitting_function}

In this appendix, we present a functional form that provides a good fit to our model cumulative SFH  (as shown in e.g. Fig.~\ref{fig:SFH_mstar}) for galaxies selected over the redshift range $6 \lesssim z \lesssim 10$. Although an exponential functional form for the instantaneous SFH is frequently adopted, we find that a bounded power law fit provides a better description of the model data. This function is given by:

\begin{equation}
y(x)=
\begin{cases}
\left[1-\left(\dfrac{x}{x_c}\right)^p\right]^q, & 0 \le x \le x_c, \\[6pt]
0, & x > x_c ,
\end{cases}
\label{eqn:sfh}
\end{equation}
where $x_c$ sets the cut-off of the growth, $p$ governs the slope of the growth of stellar mass, and $q$ controls the transition between the power-law growth phase and the onset of star formation.

\begin{figure}
    \includegraphics[width=\columnwidth]{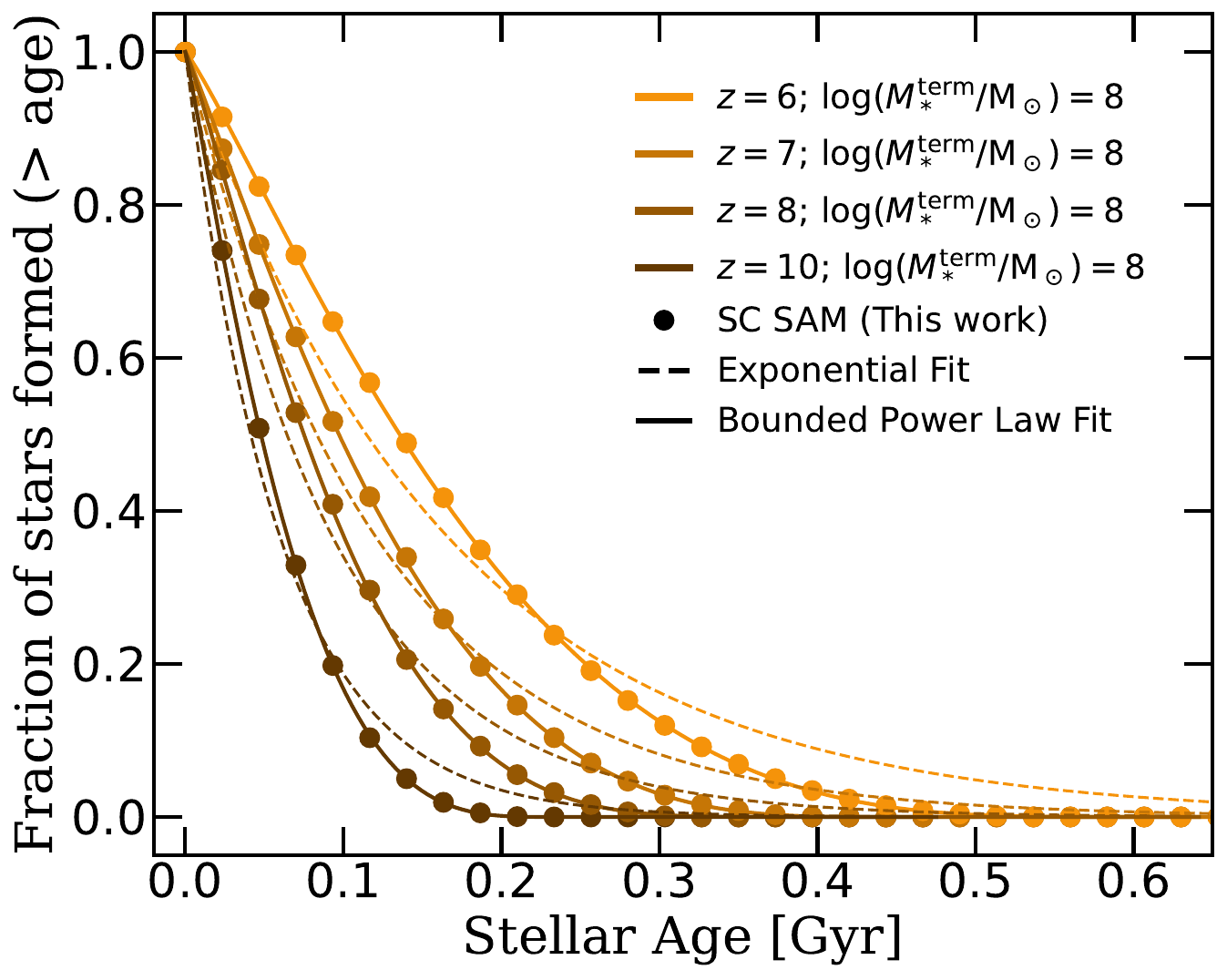}
    \caption{
        Cumulative fraction of stars formed as a function of stellar age (from old to young) for galaxies with terminal stellar masses $\log(M_*/\text{M}_\odot) = 8$ at $z=6$, 7, 8, and 10. We show that a bounded power law fit (Eqn.~\ref{eqn:sfh}) provides a better description of the model data than an exponential fit with a single e-folding timescale.
    }
    \label{fig:functional_fit}
\end{figure}


\section{Impacts of halo merger tree mass resolution on SFH}
\label{app:resolution_test}

\begin{figure}
    \includegraphics[width=\columnwidth]{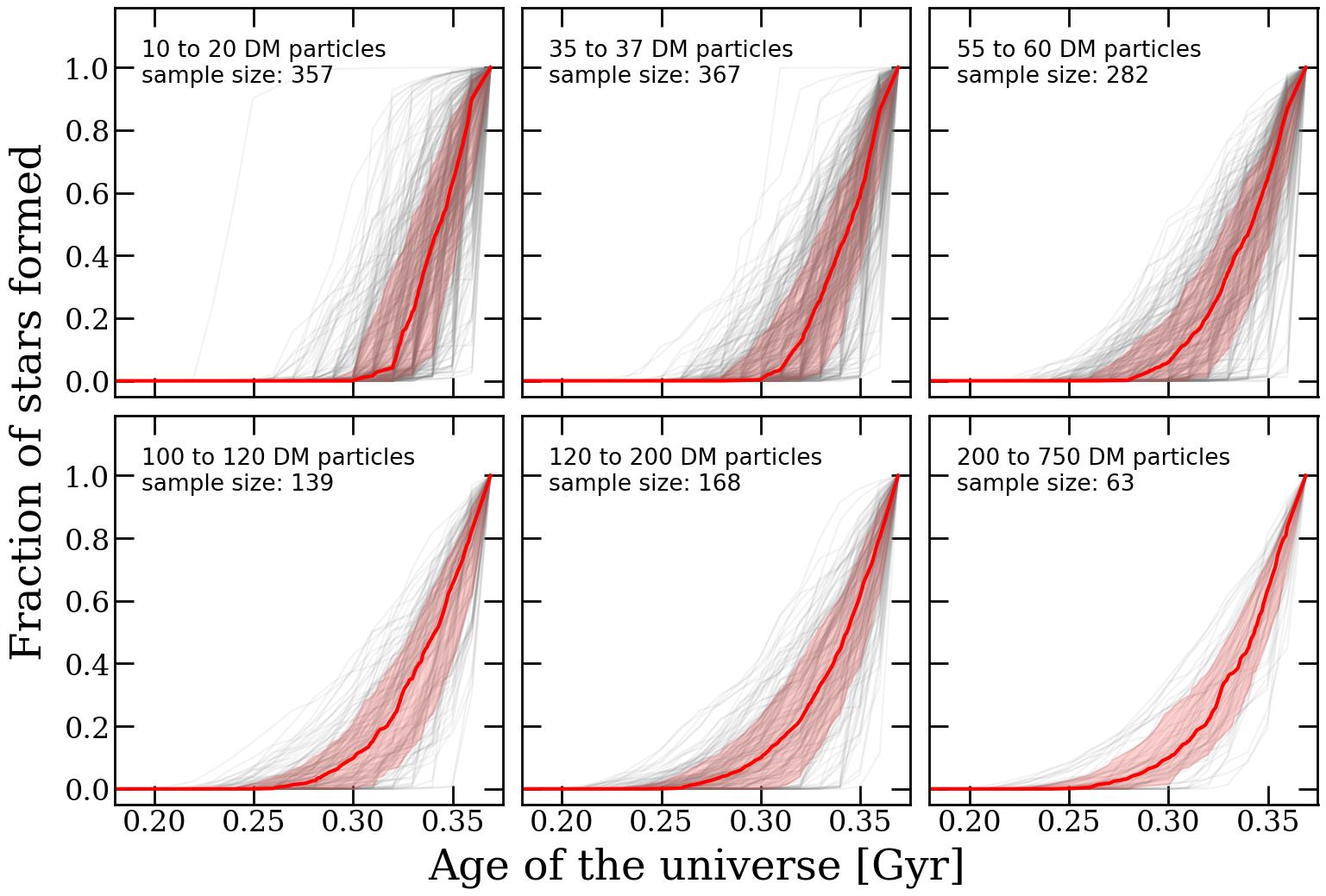}
    \caption{
        Cumulative stellar mass assembly histories normalized to the terminal stellar mass at $z=12$, shown as a function of cosmic time, for galaxies hosted by halos spanning a range of mass (expressed in terms of the number of dark matter particles) drawn from \gureft-90. Individual galaxies are shown in grey. The red solid line and shaded region indicate the median and the 16th to 84th percentile range, respectively. We find that dark matter halos with fewer than $\sim 100$ DM particles may not yield robust SFH predictions.
    }
    \label{fig:cSFH_mhalo_z12_app}
\end{figure}

\begin{figure}
    \includegraphics[width=\columnwidth]{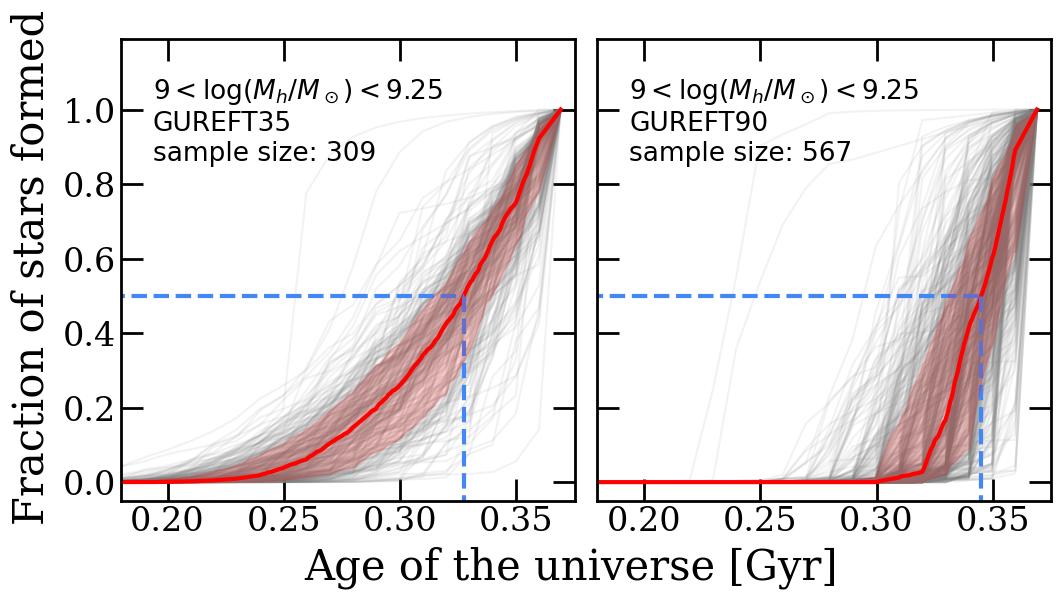}
    \caption{
        Resolution test comparing cumulative stellar mass assembly histories for galaxies hosted by halos in the mass range $9 \lesssim \log(M_\text{h}/\text{M}_{\odot}) \lesssim 9.25$ at $z=12$, drawn from \gureft-90 and \gureft-35. This halo-mass interval corresponds to 12--21 DM particles in \gureft-90 and 199--355 particles in \gureft-35, illustrating how merger-tree mass resolution affects the inferred early growth and the apparent onset of star formation. The blue dashed line marks the cosmic time at which the median assembly history reaches 50 per cent of the terminal stellar mass. This illustrates that insufficiently resolved merger trees can have a significant impact on the predicted SFHs and characteristic star formation timescales.
    }
    \label{fig:cSFH_mhalo_G35G90_app}
\end{figure}

\begin{figure*}
    \includegraphics[width=2\columnwidth]{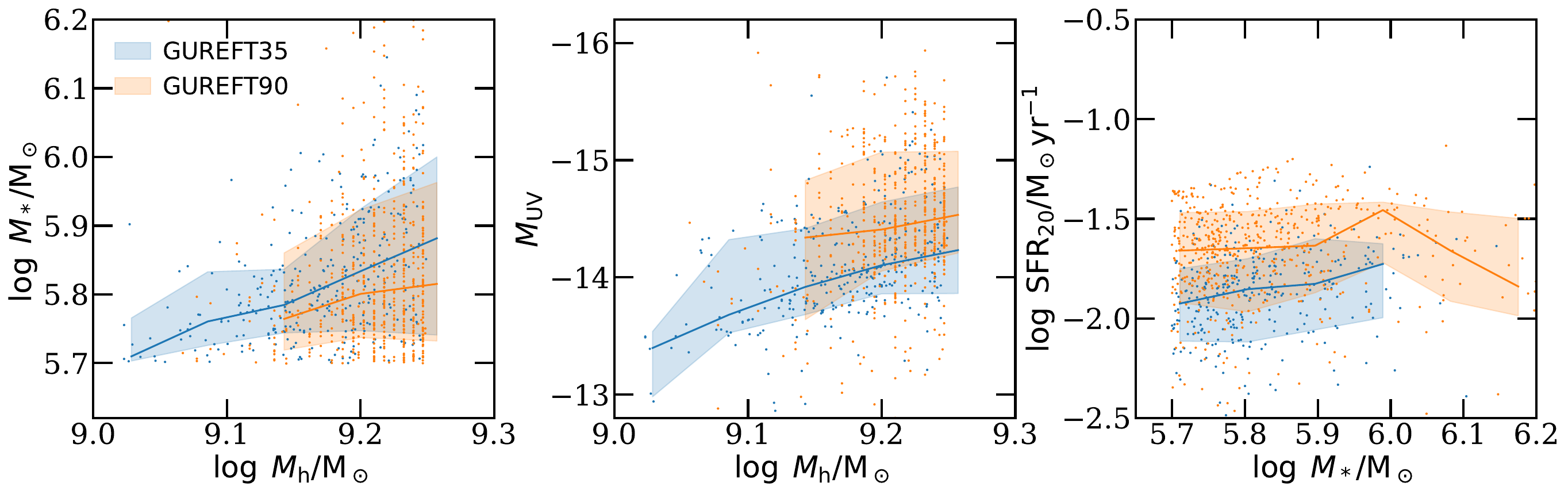}
    \caption{
        Comparison of the $M_\text{halo}$--$M_*$, $M_\text{halo}$--$M_\text{UV}$, and $M_*$--SFR$_{20}$ relations for galaxies hosted by halos in the mass range $9 \lesssim \log(M_\text{h}/\text{M}_{\odot}) \lesssim 9.25$ at $z=12$ (same as those shown in Fig.~\ref{fig:cSFH_mhalo_G35G90_app}), drawn from the \gureft-35 and \gureft-90 volumes. These comparisons highlight that, when merger trees are insufficiently resolved, the SAMs systematically underpredict $M_*$ while overpredicting $M_\text{UV}$ and short-timescale SFRs.
    }
    \label{fig:prop_res}
\end{figure*}

\begin{figure*}
    \includegraphics[width=2\columnwidth]{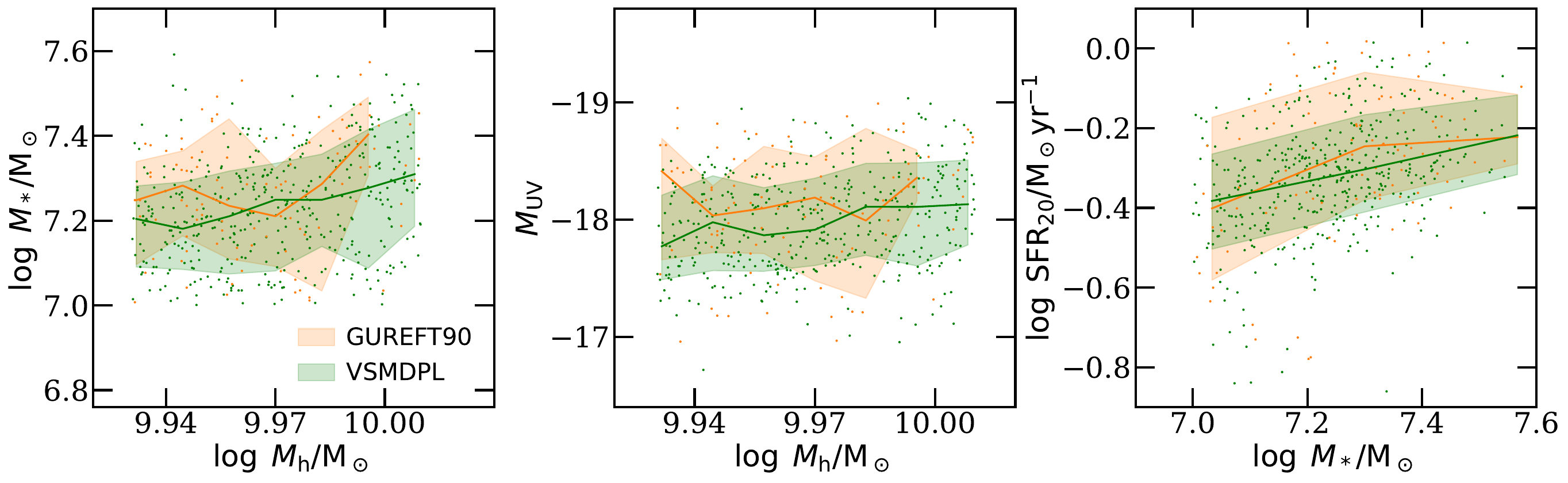}
    \caption{
        Similar to Fig.~\ref{fig:prop_res}, this figure shows a comparison of the $M_\text{halo}$--$M_*$, $M_\text{halo}$--$M_\text{UV}$, and $M_*$--SFR$_{20}$ relations for galaxies hosted by halos with 100--120 DM particles at $z\sim12$ drawn from \gureft-90 ($9.93 \lesssim \log(M_\text{h}/\text{M}_{\odot}) \lesssim 10.01$) and those falling within the same mass range drawn from VSMDPL, resolved with 932--1119 DM particles. This shows that the predicted galaxy properties mostly converge when merger trees are sufficiently well resolved.
    }
    \label{fig:prop_res_vg}
\end{figure*}

Dark matter halo merger trees provide the backbone for semi-analytic models of galaxy formation, and it is well known that a sufficiently large number of particles is required to robustly resolve halos, especially their structural properties and related quantities (see discussion in \citealt{Yung2024a} and references therein). Similarly, particle resolution can also affect inferred merger histories, where halos near the mass-resolution limit contain only a small number of particles, and their merger trees can be truncated prematurely.

In this appendix, we quantify how mass resolution impacts the predicted star formation histories. In Fig.~\ref{fig:cSFH_mhalo_z12_app}, we show cumulative stellar-mass assembly histories (normalized to the terminal stellar mass at $z=12$) for galaxies hosted by haloes drawn from \gureft-90, organized by the number of dark matter particles in their host haloes, spanning from a few hundred particles down to only a few tens of particles. It is noteworthy that these assembly histories become systematically more \textit{choppy} and \textit{discrete} as the number of particles decreases. Galaxies simulated within such poorly resolved merger trees exhibit larger scatter and substantially less reliable early-time SFHs.

To further investigate the impact of poorly resolved merger trees on SFHs, in Fig.~\ref{fig:cSFH_mhalo_G35G90_app}, we compare cumulative stellar mass assembly histories (normalized to the terminal stellar mass at $z=12$) for galaxies hosted by halos in the interval $9 < \log(M_\text{h}/\text{M}_\odot) < 9.25$. This range corresponds to 12--21 dark matter particles in \gureft-90 and 199--355 dark matter particles in \gureft-35. The comparison highlights that poorly resolved halos can yield systematically different early-time growth behaviour, reflecting limitations in resolving the earliest progenitor stages.

In Fig.~\ref{fig:prop_res}, we compare the $M_\text{halo}$--$M_*$, $M_\text{halo}$--$M_\text{UV}$, and $M_*$--SFR$_{20}$ relations for galaxies hosted by halos in the same mass range as shown in Fig.~\ref{fig:cSFH_mhalo_G35G90_app}. These comparisons illustrate that merger trees with insufficient mass resolution (e.g. those sourced from \gureft-90) would yield systematically under-predicted $M_*$, likely due to the later onset (or premature truncation) of star formation, as well as an over-predicted $M_\text{UV}$ and SFRs averaged over a short timescale due to violent `artificial bursts' induced by the merger of coarse progenitors.

Similarly, in Fig.~\ref{fig:prop_res_vg}, we compare halo merger trees that are deemed well-resolved with 100--120 DM particles, drawn from the \gureft-90 volume, spanning the mass range of $9.93 \lesssim \log(M_\text{h}/\text{M}_{\odot}) \lesssim 10.01$, to halos of the same mass range drawn from VSMDPL, resolved with 932--1119 DM particles. It is shown that the predicted galaxy properties converges.

Based on these tests, we recommend treating merger trees for halos resolved with fewer than $\sim$100 dark matter particles with caution, and excluding them from analyses that depend sensitively on early-time SFHs or the onset of star formation.


\section{Comparison with MD14 calculation of ${\cal{K}}_\text{UV}$}
\label{app:KUV_MD14}
\begin{figure}
	\includegraphics[width=\columnwidth]{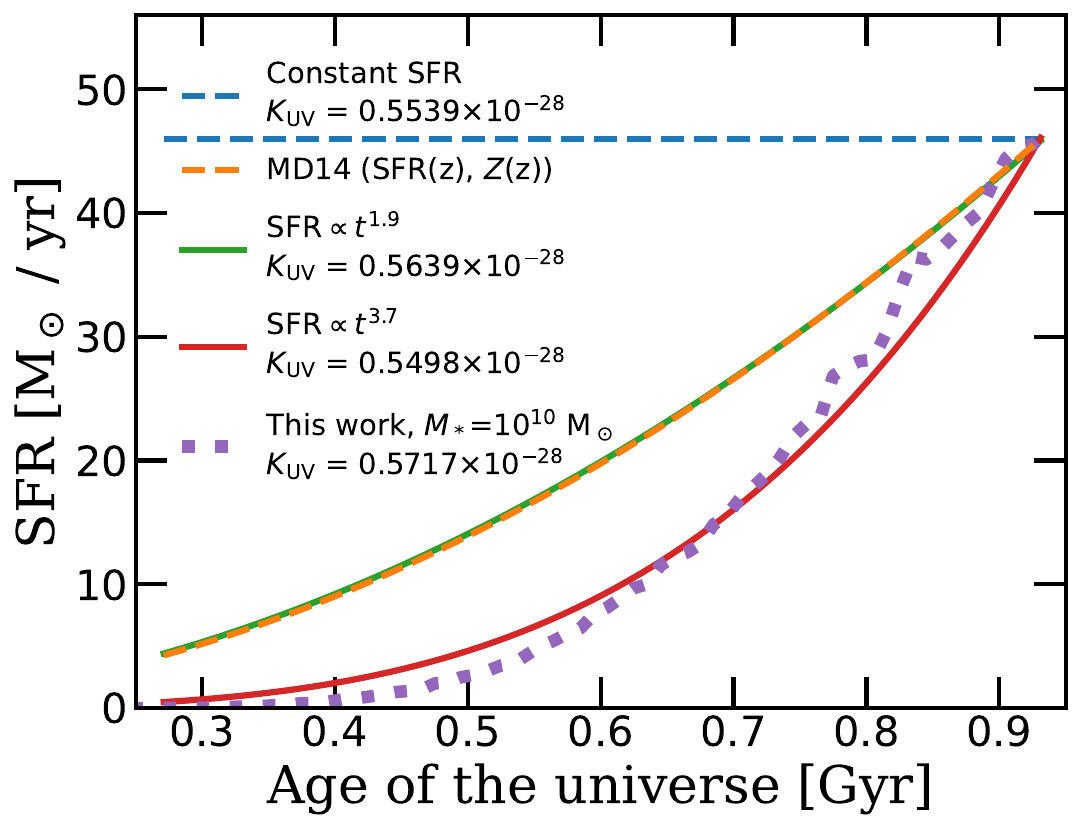}
	\caption{
		The orange dashed line shows the SFH adopted by \citetalias{Madau2014} to derive their fiducial \KUV\ (Eqn.~\ref{eqn:sf_hist}). The green solid line shows a power-law SFH, $\mathrm{SFR}\propto t^{1.9}$, which provides an excellent approximation to the \citetalias{Madau2014} SFH over $6<z<15$. The purple dotted line shows the median SFH of galaxies with $M_*=10^{10}$ \Msun\ at $z=6$, while the red solid line shows a steeper power-law SFH, $\mathrm{SFR}\propto t^{3.7}$. The \KUV\ value computed from each SFH is reported alongside the corresponding curve in the legend, in units of \Msun\ yr$^{-1}$\,/\,erg s$^{-1}$ Hz$^{-1}$.
	}
	\label{fig:KUV_comp}
\end{figure}

To compute the fiducial value of ${\cal{K}}_\text{UV}=1.15 \times 10^{-28}\, M_{\odot}$ \ yr$^{-1}$\,/\,erg s$^{-1}$ Hz$^{-1}$, \citetalias{Madau2014} adopt a star formation history following their Eqn.~15:
\begin{equation}
\label{eqn:sf_hist}
{\rm SFR}(z) \propto \frac{(1+z)^{2.7}}{1 + [(1+z)/2.9]^{5.6}}
\end{equation}
and a metallicity history (i.e. mean stellar metallicity as a function of redshift) given by
\begin{equation}
\label{eqn:metallicity_hist}
Z_*(z) = Z_\odot\, 10^{-0.15z}
\end{equation}
They assume that the onset of star formation is at $z=12$.

It is worth noting that Eqn.~\ref{eqn:sf_hist} (Eqn.~15 in \citetalias{Madau2014}) represents the \emph{cosmic star formation history of all galaxies}, and was derived from observations of $z\lesssim 8$ galaxies. The \citetalias{Madau2014} approach to computing \KUV\ 1) assumes that the typical SFH \emph{of every galaxy} is well represented by this cosmic star formation history (with normalization scaled up or down) and 2) extrapolates the cosmic SFH measured at $z\lesssim 8$ to the redshift regime $8<z<12$. The adopted metallicity evolution introduces an additional modelling assumption that remains comparatively uncertain at these redshifts.

In Fig.~\ref{fig:KUV_comp}, we show the SFH adopted by \citetalias{Madau2014} to derive their fiducial \KUV, and compare this with the median SFH of our more massive model galaxies with $M_* = 10^{10}$ \Msun\ at $z=6$ (the same population shown in Fig.~\ref{fig:SFH_mstar}). Both the \citetalias{Madau2014} SFH and our model-predicted median SFH are well represented by simple power laws in cosmic time. Over the redshift range $6 < z < 15$, the \citetalias{Madau2014} SFH is well approximated by SFR $\propto t^{1.9}$, while the SAM galaxy SFH is well described by a steeper power law with SFR $\propto t^{3.7}$. Thus although the SFH adopted by \citetalias{Madau2014} may not seem extremely well motivated, we can see that their adopted SFH turns out to be quite similar to the ones the emerge from our physics-based models. 

We report the resulting \KUV\ values for a constant SFR, for the \citetalias{Madau2014} SFH and for the SAM-motivated composite median SFH in Fig.~\ref{fig:KUV_comp} and find that \KUV\ is very insensitive to the range of SFHs considered here. This is because, as seen in Fig.~\ref{fig:MUV_age}, the contribution to the far UV light drops sharply for ages greater than 20-30 Myr. For a power law SFH with the adopted range of slopes, the SFR only drops by about 10\% over this timescale, and most of the change in luminosity resulting from this drop (relative to a constant SFR) is cancelled out by the correspondingly lower average SFR. 
All cases shown here adopt the redshift-dependent metallicity history used by \citetalias{Madau2014}, as given by Eqn.~\ref{eqn:metallicity_hist}.
We note that \KUV\ remains sensitive to the overall stellar metallicity, since metallicity systematically shifts $L_{\rm UV}$ across all stellar ages (Fig.~\ref{fig:MUV_age}). However, for the cases examined here, the resulting \KUV\ values are largely insensitive to whether we adopt a redshift-dependent metallicity history or a fixed stellar metallicity.

The value of \KUV\ is also sensitive to the IMF and its upper mass cutoff, and the SSP models adopted. We show the values of \KUV\ that we have computed, adopting the MD14 SFH and $Z(z)$ for several different cases in Table~\ref{tab:KUV}.

\begin{table}[h]
\centering
\caption{\KUV\ values}
\label{tab:KUV}
\begin{tabular}{|l|c|c|c}
\hline
SSP model & IMF & upper cutoff & \KUV\ \\
&& (\Msun) &($10^{-28}$ (\Msun/yr)/(erg/s/Hz) \\ \hline
FSPS  & Salpeter & 100 &  1.00         \\ 
FSPS  & Chabrier &  100 &   0.635         \\ 
FSPS  &  Chabrier & 300 &    0.638        \\
BPASS & Chabrier & 300 & 0.564 \\
\hline
\end{tabular}
\end{table}

\section{Updated figures from Yung et al. 2024a}

\begin{figure*}
	\includegraphics[width=1.35\columnwidth]{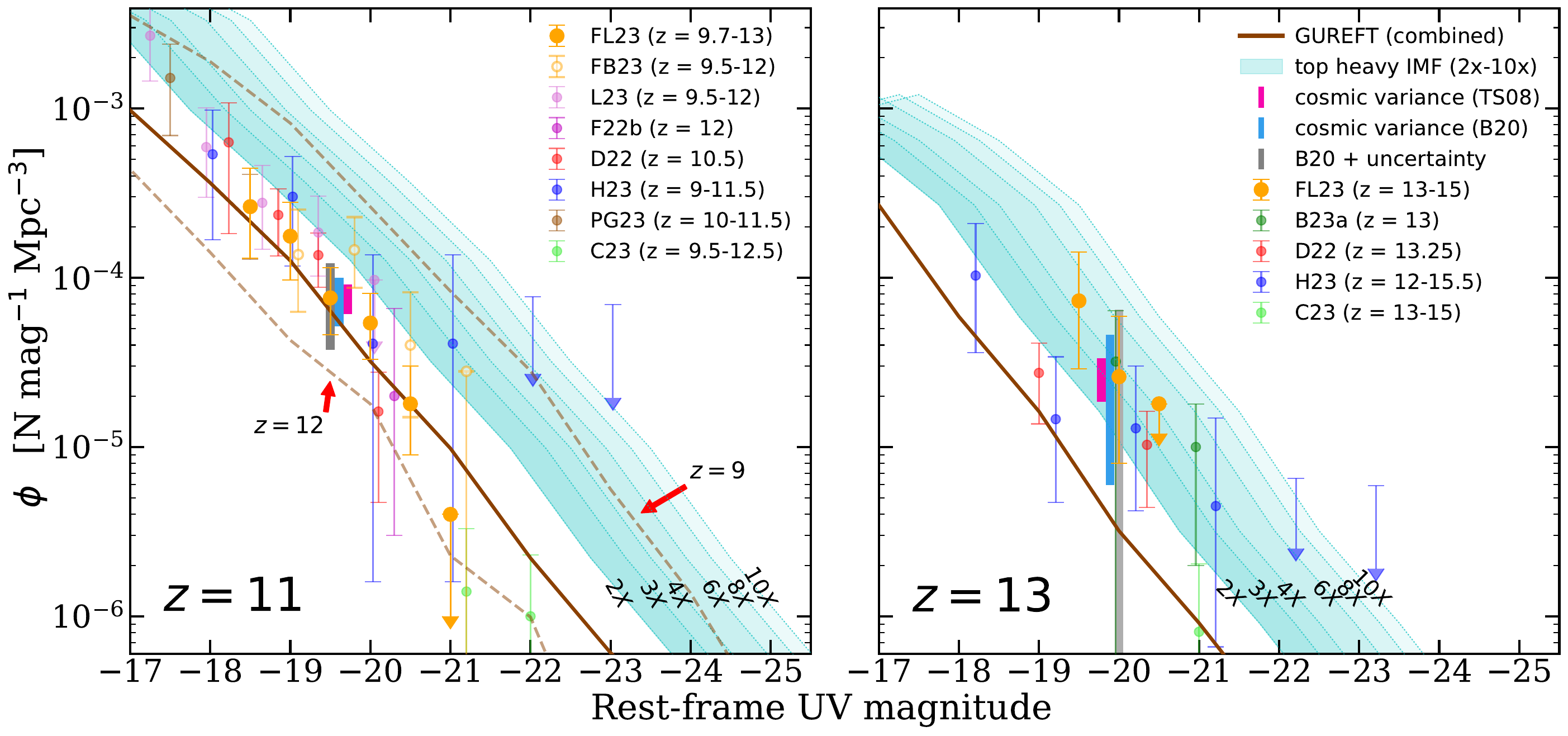}
	\caption{
		Updated version of Fig.~4 from \citetalias{Yung2024}, providing a more detailed look at the UV LFs at $z = 11$ (left) and $z = 13$ (right) . The combined outputs across the four GUREFT boxes are shown by the brown line. In the left panel, in addition to the $z = 11$ predictions, we also show the predicted UV LFs at $z = 9$ and 12 with the light brown dashed line above and below, respectively, as the observational sample of F23 spans a redshift range of $z = 9.5$ -- 12. In addition, we show two estimates of the 1$\sigma$ uncertainty on the number density due to cosmic variance with the blue and pink error bars (see the text for a full description of how these were calculated). The grey bar shows a representative total error, including the quoted error bar from the observational study combined with the estimated cosmic variance in quadrature. The light blue shaded regions and lines show the effect of `boosting' the UV luminosity of all galaxies in our model by a factor of 2--10, as indicated on the plot label. This illustrates that a boosted UV light-to-mass ratio \textbf{is no longer needed} to bring our predictions into agreement with the observational measurements at $z\sim 11$.
	}
	\label{fig:gureft_UVLF_var}
\end{figure*}

\begin{figure*}
	\includegraphics[width=1.75\columnwidth]{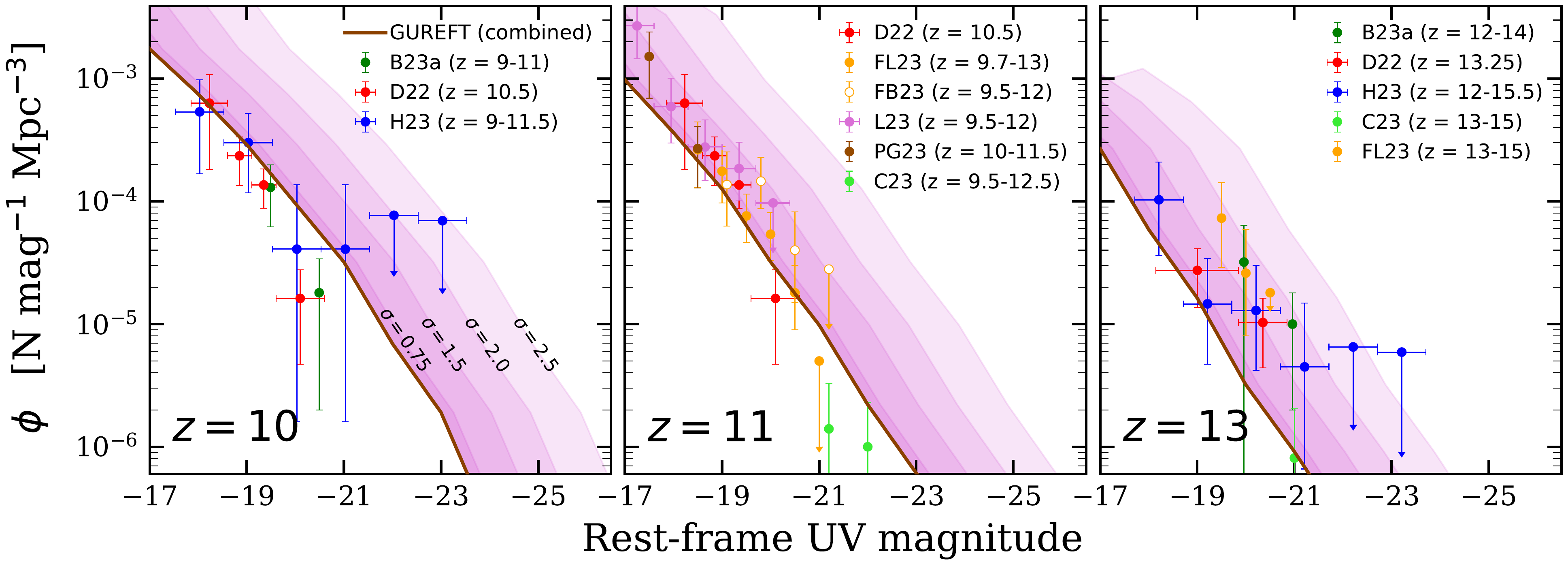}
	\caption{
		 Updated version of Fig.~5 from \citetalias{Yung2024}, providing a more detailed look at the UV LFs at z = 10 (left), z = 11 (middle), and z = 13 (right), where the fiducial predictions from this work made with the GUREFT merger trees are shown in brown, observational compilations as shown in the corresponding panels in Fig.~\ref{fig:gureft_UVLF_var}, and the potential effects of UV stochasticity are shown with different shades of pink as labelled on the plot panels. This effect is implemented by adding a stochastic component to the predicted UV magnitudes of individual galaxies in post-processing, assuming the random component follows a Gaussian distribution centred at zero with standard deviation $\sigma$. We find that bursty star formation with a scatter of $\sigma 1.5$--2 \textbf{is no longer needed} to bring our fiducial model into agreement with observations at $z \lesssim 11$.
	}
	\label{fig:gureft_UVLF_burst}
\end{figure*}

\citetalias{Yung2024} provided a detailed, timely assessment of the early \textit{JWST} results on high- to ultra-high-redshift galaxy candidates \citep[e.g.][]{Finkelstein2022, Finkelstein2023, Castellano2022, Donnan2022, Adams2023}, and contributed to the discussion regarding their number densities and physical origins \citep[e.g.][]{Naidu2022, Boylan-Kolchin2023, Labbe2023, Robertson2023}. In particular, \citetalias{Yung2024} quantified several sources of uncertainty affecting both observation-reported and model-predicted UV LFs, clarifying how these uncertainties can play a part in explaining the the apparent discrepancy between observations and theory. On the observational side, these include field-to-field variance due to the limited survey areas of deep extragalactic surveys, as well as redshift uncertainties associated with photometric redshifts. On the modelling side, these include uncertainties in the light-to-mass ratio (e.g., from plausible IMF variations) and the impact of short-timescale star formation variability on UV luminosities. 

While results in \citetalias{Yung2024} that depend primarily on halo demographics and stellar masses are unaffected by the updates introduced in this work, it is important to note that the finer stellar age binning in our improved SFH$\rightarrow$SED routine have yielded significant shifts in the predicted UV LFs. This, in turn, affects some of the interpretation associated with figs.~4 and 5 in \citetalias{Yung2024}, which explored the degree of UV `boosting' or UV stochasticity required to reconcile model predictions with observations. In this Appendix, we reproduce those figures using the updated UVLFs from this work. Observational constraints and data labels in these plots are preserved, including \citet[][D22]{Donnan2022}, \citet[][H23]{Harikane2023}, \citet[][CEERS Epoch 1 only; hereafter FB23]{Finkelstein2023}, \citet[][full CEERS field; hereafter FL23]{Finkelstein2023}, \citet[][F22a]{Finkelstein2022}, \citet[][B23a]{Bouwens2023}, \citet[][PG23]{Perez-Gonzalez2023}, \citet[][L23]{Leung2023a}, and \citet[][C23]{Casey2024}.


\bsp	
\label{lastpage}
\end{document}